\documentclass[aps,prd,12pt,nofootinbib]{revtex4}

\usepackage{comment}

\usepackage{epsfig}
\usepackage{graphicx}
\usepackage[font=small,skip=0pt,justification=raggedright]{caption}
\usepackage{amsmath}
\usepackage{amssymb}
\usepackage{mathrsfs}
\usepackage{verbatim}
\usepackage{dutchcal}

\usepackage[normalem]{ulem}
\usepackage{xcolor}

\newcounter{fig}   \newcommand{\lbfig}[1]{\refstepcounter{fig}
\label{#1} }

\newcommand{\bea}{\begin{eqnarray}}
\newcommand{\eea}{\end{eqnarray}}
\newcommand{\be}{\begin{equation}}
\newcommand{\ee}{\end{equation}}

\newcommand{\re}[1]{(\ref{#1})}

\newcommand{\cgg}{\mathcal{g}}

\newcommand{\eqn}{\begin{eqnarray}}
\newcommand{\eqnx}{\end{eqnarray}}

\date{\today}

\begin{document}
\title{
Charged hairy black holes in the gauged Einstein-Friedberg-Lee-Sirlin model}
\author{J.~Kunz}
\affiliation{Institute of Physics, University of Oldenburg,
Oldenburg D-26111, Germany}
\author{Ya.~Shnir}
\affiliation{BLTP, JINR, Dubna 141980, Moscow Region, Russia\\
Institute of Physics,
Carl von Ossietzky University Oldenburg, 
Oldenburg D-26111, Germany
}

\begin{abstract}
We obtain charged spherically symmetric black holes in the two-component scalar Einstein-Maxwell-Friedberg-Lee-Sirlin model with a symmetry breaking potential.  
These asymptotically flat black holes carry resonant scalar Q-hair. As expected, these hairy black holes give rise to non-uniqueness.
When comparing these solutions with the corresponding charged boson stars and Reissner-Nordstr\"om black holes, we find a different pattern in the case of a massive real scalar component and a massless one.
We demonstrate that, as the real component becomes massless, the resonant hairy black holes bifurcate from Reissner-Nordstr\"om black holes for sufficiently small gravitational coupling.

\end{abstract}
\maketitle

\section{Introduction}
One of the interesting recent developments in General Relativity (GR) is associated with the discovery of hairy black holes (see e.g.~\cite{Volkov:1998cc,Herdeiro:2015waa}).
In fact, there are various ways known to circumvent the well-known no-hair theorems in GR.
One way would be to involve stationary non-Abelian fields.
Then the non-linearity of the matter fields allows for black holes with non-Abelian hair \cite{Luckock:1986tr,Volkov:1989fi,Bizon:1990sr,Droz:1991cx}. 
Another type of hairy black holes arises for rotating configurations with matter fields that possess a harmonic time-dependence when the synchronization condition is imposed, that represents the threshold of superradiance \cite{Hod:2012px,Herdeiro:2014goa}.
Such synchronized hair is carried by black holes of the Einstein-Klein-Gordon system:
These black holes with scalar hair are continuously connected to the Kerr black hole \cite{Hod:2012px,Herdeiro:2014goa,Benone:2014ssa,Kunz:2019bhm}. 
Similarly, synchronized hairy black holes exist in models with time-dependent non-Abelian matter fields \cite{Herdeiro:2018djx,Herdeiro:2018daq}.
In all these cases the hairy generalizations of the vacuum Kerr black holes bifurcate from the Kerr black holes at the threshold of superradiance \cite{Brito:2015oca}. In particular, they arise from the presence of linear scalar clouds, when backreaction is taken into account.

A somewhat different mechanism is at work for hairy Reissner-Nordstr{\"o}m (RN) black holes. 
In this case there are no linear clouds of a charged complex scalar field at the threshold of superradiance \cite{Hod:2012wmy,Hod:2013nn} although quantum superradiant modes exist for a massless charged scalar field on an RN 
space-time \cite{Balakumar:2022yvx,Balakumar:2023trj}.
However, when the scalar field is self-interacting, charged non-linear scalar Q-clouds around RN black holes may exist \cite{Herdeiro:2020xmb,Hong:2020miv}.
The scalar field then needs to satisfy the so-called superradiance resonance condition at the horizon.
When backreaction is taken into account, these cloudy black holes turn into spherically symmetric charged black holes with scalar hair \cite{Herdeiro:2020xmb,Hong:2020miv}.
Since the Q-hairy black holes do not emerge from linear clouds they feature a gap with respect to RN black holes.
On the other hand, these Q-hairy black holes are connected to boson stars \cite{Kaup:1968zz,Feinblum:1968nwc,Ruffini:1969qy,Deppert:1979au,Colpi:1986ye,Friedberg:1986tp,Friedberg:1986tq,Jetzer:1989av,Jetzer:1989us,Jetzer:1992tog,Kleihaus:2005me,Kleihaus:2007vk,Pugliese:2013gsa,Collodel:2017biu,Herdeiro:2020kvf,Herdeiro:2021mol,Kunz:2021mbm}. 
When the black hole horizon radius goes to zero, the global charges smoothly reach the corresponding values of the solitonic boson stars.

In the present paper we extend our recent study of  charged spherically symmetric boson stars in the Einstein-Maxwell-Friedberg-Lee-Sirlin (EMFSL) model \cite{Kunz:2021mbm} by including a small charged black hole at their center.
As in the case of Q-hairy black holes with self-interacting scalar fields, the resonance condition must be satisfied at the horizon for these new Q-hairy black holes.
Here the role of the higher (sextic) order self-interaction potential is, however, played by the interacting two-component scalar fields, that allow these hairy black holes to exist. 

The key difference between the Q-hairy black holes in the two-component EMFLS model and similar solutions with resonant single-component scalar hair is that the real component of the EMFLS model can be long-ranged \cite{Levin:2010gp,Loiko:2018mhb}. 
The corresponding Coulomb-like asymptotic force decays in the same way as the attractive gravitational potential and the long-range electrostatic interaction. 
The corresponding balance of competing forces may then open a window for the possible existence of linear clouds around RN black holes, similar to those around the Kerr black holes.

We here investigate the properties of these Q-hairy black holes and determine their domain of existence.
Similar to EMFSL boson stars, these properties depend distinctly on the mass of the real scalar component.
In particular, the domain of existence is very different for vanishing or finite mass.
Our results indicate that, as the real scalar component becomes massless, the resonant Q-hairy EMFLS black holes bifurcate from the RN black holes when the gravitational coupling constant is sufficiently small.

This paper is organized as follows. In Section II we present the EMFLS model, the field equations and the stress-energy tensor of the system of interacting fields. 
Here we also discuss the gauge fixing, the physical quantities of interest and present the spherically symmetric parametrization of the metric and the matter fields. 
In Section III we present the results obtained by solving the coupled system of equations numerically. Here we study the dependence of the resonant Q-hairy EMFLS black holes on the strength of the gravitational coupling constant as well as the dependence on the horizon area. 
Then we address the limit when the mass of the real scalar field vanishes such that it becomes long-ranged. We close with our conclusions in Section IV.

\section{The model}

The Einstein-Maxwell-Friedberg-Lee-Sirlin model \cite{Kunz:2021mbm,Friedberg:1976me} describes a self-gravitating coupled system of a $U(1)$ gauged (3+1)-dimensional complex scalar field $\phi$, minimally interacting with an Abelian vector potential $A_\mu$ and a real self-interacting scalar field $\psi$.
The corresponding action is given by \footnote{
We here employ the same notation as in Refs.~\cite{Kunz:2019sgn,Kunz:2021mbm}.}
\be S=\int d^4 x \sqrt{-\cgg} \left(\frac{R}{4\alpha^2} + L_{m}\right),
\label{lag}
\ee
where the matter field Lagrangian is
\be
L_{m} = -\frac14 F_{\mu\nu}F^{\mu\nu} - D_\mu\phi^*D^\mu\phi -\partial_\mu\psi \partial^\mu\psi -
m^2 \psi^2|\phi|^2 - \mu^2 (\psi^2 -v^2)^2 \, .
\label{GaugedFLS}
\ee
Here $R$ is the Ricci scalar associated with the spacetime metric $g_{\mu\nu}$ with the determinant $\cgg$ and $\alpha^2=4\pi G$ is the effective gravitational coupling with Newton's constant $G$.
The matter field Lagrangian contains the $U(1)$ field strength tensor $F_{\mu\nu}=\partial_\mu A_\nu-\partial_\nu A_\mu$, the covariant derivative of the complex scalar field $\phi$, $D_\mu\phi = \partial_\mu \phi - ig A_\mu\phi$ with gauge coupling $g$, the kinetic term of the real scalar field $\psi$, as well as the symmetry breaking scalar field potential of the model \cite{Friedberg:1976me}
\be
U(\psi,\phi)=m^2 \psi^2 |\phi|^2 + \mu^2(\psi^2-v^2)^2   \, 
\label{pot}
\ee
with positive constants $m$, $\mu$ and $v$.

The global minimum of the potential $U(\psi,\phi)$ corresponds to $\psi=v$ and $|\phi|=0$, where the fields assume their vacuum expectation values. 
In the vacuum $D_\mu \phi =0$, $\partial_\mu \psi =0$, and $F_{\mu\nu}=0$.
The mass of the scalar excitations is defined in terms of the parameters of the potential. 
The mass of the real scalar excitations of $\psi$ is given by $m_\psi= \sqrt{ 8} \mu v$ and the complex scalar component $\phi$ acquires mass due to the coupling with the real partner $\psi$, $m_\phi = mv$.
Note that the EMFLS model \re{lag} reduces to the Einstein-Maxwell-Klein-Gordon theory as $\mu \to \infty$ and the real component decouples, $\psi=v$. 
Another limit corresponds to $\mu \to 0$, where the real scalar field $\psi$ becomes long-ranged.
In vacuum ($|\phi|=0$) the mass of the gauge excitations $A_\mu$ is zero and the gauge field is long-ranged. 
Inside the configurations, however, the gauge field acquires a mass due to the coupling with the complex scalar field $\phi$. 

The EMFLS model \re{lag} is invariant with respect to local $U(1)$ transformations
\be
\phi \to \phi e^{ig \xi(x)},\quad A_\mu \to A_\mu + \partial_\mu \xi (x)\, ,
\ee
where the associated conserved Noether current is given by
\be
j_\nu = i( D_\nu \phi^* \, \phi - \phi^* D_\nu \phi ) \, .
\label{current}
\ee
Variation of the action \re{lag} with respect to the metric and the gauge potential leads to the Einstein-Maxwell equations
\be
\begin{split}
R_{\mu\nu} -\frac12 R g_{\mu\nu} &= 8\pi G\left( T_{\mu\nu}^{Em} + T_{\mu\nu}^{Sc}\right)\, ,\\
\partial_\mu(\sqrt{-\cal g} F^{\mu\nu})& = g \sqrt{-\cal g} j^\nu\, , \\
\end{split}
\label{field-eqs-em}
\ee
where the source term in the Maxwell equations is the conserved Noether current \re{current}.
The two components of the stress-energy tensor of the electromagnetic and the scalar fields on the right hand side of the Einstein equations are given by
\be
\begin{split}
T_{\mu\nu}^{Em} &=F_\mu^\rho F_{\nu\rho} - \frac14 g_{\mu\nu} F_{\rho\sigma}  F^{\rho\sigma}\,,\\
T_{\mu\nu}^{Sc} &=D_\mu\phi^* D_\nu\phi + D_\nu\phi^* D_\mu\phi +  \partial_\mu\psi
\partial_\nu\psi\\
& - g_{\mu\nu}\left(\frac{g^{\rho\sigma}}{2} (D_\rho\phi^* D_\sigma\phi +
D_\sigma\phi^* D_\rho\phi +  \partial_\rho\psi
\partial_\sigma\psi) +  U(\phi,\psi)\right) \, .
\end{split}
\label{Teng}
\ee
Variation with respect to the scalar fields leads to the scalar equations 
\be
\begin{split}
\partial^\mu \partial_\mu\psi & = 2\psi (m^2 |\phi|^2  +2 \mu^2 (1-\psi^2)) \, , \\
D^\mu D_\mu\phi & = m^2 \psi^2 \phi \, .
\end{split}
\label{field-eqs-scalar}
\ee
Rescaling of the fields and the radial coordinate allows to set $v=1$ and $m=1$, leaving us with dependence on three model parameters, $g$, $\mu$ and $\alpha$ \cite{Kunz:2019sgn,Kunz:2021mbm}.

\subsection{The Ansatz}

To obtain spherically symmetric solutions we choose for the metric the Ansatz
\be
ds^2=g_{\mu\nu}dx^\mu dx^\nu= -F_0(r)dt^2 +F_1(r)(dr^2 + r^2 d\Omega^2 )
\label{metric}
\ee
where $d\Omega^2= d\theta^2 + \sin^2\theta d\varphi^2$, and the metric functions $F_0$ and $F_1$ depend on the isotropic radial coordinate $r$ only. 
By a reparametrization of the metric this Ansatz is readily transformed to Schwarzschild-like coordinates.
We here prefer to work with the line element \re{metric} which can be directly generalized to an axially symmetric form \cite{Kunz:2019bhm,Herdeiro:2018djx}. 

Imposing spherical symmetry and a vanishing magnetic field we employ for the scalar field the Ansatz 
\be
\label{scalans}
\psi=X(r)\, , \qquad  \phi=Y(r)e^{i\omega t}\, ,
\ee
where $\omega$ is the angular frequency of the complex scalar field $\phi$,
and for the gauge potential
\be
\label{Aans}
A_{\mu} dx^{\mu} =A_0(r)dt \, .
\ee

We note, that the $U(1)$ invariance of the EMFLS model admits the unitary gauge fixing, ${\rm Im}~\phi=0$ \cite{Herdeiro:2020xmb,Kleihaus:2009kr,Loginov:2020xoj,
Brihaye:2020vce,Brihaye:2021phs,Brihaye:2021mqk}.
However, as in our previous study of the spherically symmetric boson stars of the EMFLS model \cite{Kunz:2021mbm}, we employ for the hairy black holes the static gauge fixing $A_0(\infty)=0$ and retain the dependence of the solutions on the angular frequency $\omega$. 
Then the only non-vanishing component of the Noether current \re{current} is $j_0$.

Our goal is to find solutions of the EMFLS model with a static, topologically spherical event horizon, located at a constant value of radial variable $r=r_H>0$ in an asymptotically flat spacetime. 
The condition of finiteness of the energy-momentum tensor \re{Teng} and the charge density \re{charge} on the event horizon implies the resonance condition between the scalar angular frequency $\omega$ and the value of the electric potential $A_0$ \cite{Herdeiro:2020xmb,Hong:2020miv}
\be
\omega = g A_0(r_H) \, ,
\label{synchro}
\ee
which needs to be imposed as a boundary condition on the function $A_0$ on the horizon.

It is convenient to make use of the following exponential parametrization for the metric functions
\be
F_0(r)=\frac{\left(1-\frac{r_H}{r} \right)^2}{\left(1+\frac{r_H}{r} \right)^2} e^{2f_0(r)};\quad
F_1(r)=\left(1+\frac{r_H}{r} \right)^4e^{2f_1(r)}\, .
\label{metric-hor}
\ee
The full system of field equations \re{field-eqs-em},\re{field-eqs-scalar} is then solved numerically subject to the boundary conditions
\begin{itemize}
    \item at $r=r_H:\qquad \partial_r X = \partial_r Y = \partial_r f_0(r)=\partial_r f_1(r) = 0$ \ ,
    \item at $r=\infty:\qquad X=1,\quad Y=f_0(r)=f_1(r)=0$ \, ,
\end{itemize}
which follow from requiring regularity of the fields on the horizon and asymptotic flatness of the metric together with the appropriate vacuum state at spatial infinity.
In addition, the synchronization condition \re{synchro} is imposed, while the above gauge fixing condition also yields $A_0(r) \to 0$ for $r\to \infty$.

\subsection{Physical quantities}

Asymptotic expansions of the matter fields and the metric functions  at the horizon and at spatial infinity yield a number of important physical observables. 
The ADM mass $M$ can be read from the asymptotic behaviour of the metric function $g_{00}$ 
\be
g_{00}(r) \xrightarrow[r\to \infty]{}  -1 + \frac{\alpha^2 M}{\pi r} + O(r^{-2}) \, ,
\ee
and the total electric charge $Q$ from the gauge potential 
\be
A_0(r)\xrightarrow[r\to \infty]{}  \frac{Q}{r} + O(r^{-2}) \, .
\ee

The ADM mass can be represented as the sum of the contributions from the event horizon and the electromagnetic and scalar fields outside the horizon, 
\be
M = M_H +  M_V = -\frac{1}{2\alpha^2}\oint\limits_{\Sigma} d\Sigma_{\mu\nu}\nabla^\mu \xi^\nu -
\frac{1}{\alpha^2}\int\limits_V d\Sigma_\mu (2T_\nu^\mu \xi ^\nu - T\xi^\mu) \, ,
\ee
with Killing vector field $\xi = \partial_t$.
The total electric charge of the configuration is the sum of the electric 
horizon charge $Q_H$ and the Noether charge $Q_N$ outside the horizon 
\be
Q= Q_H+ g\, Q_N \, 
\label{Q_Q_H}
\ee
with horizon charge
\be
Q_H = \frac{1}{4\pi}\oint\limits_{\Sigma} d \Sigma_r F^{0r} =
r_H^2 \partial_r A_0 e^{-f_0-f_1} \, ,
\label{QH}
\ee
here $\Sigma$ is the horizon 2-sphere, and Noether charge
\be
Q_N=\int d^3 x \, \sqrt{-g} \, j^0 = 8\pi \int_{r_H}^\infty dr\, r^2 \frac{F_1^{3/2}}{\sqrt{F_0 }}
 (\omega - g A_0) Y^2 \, .
\label{charge}
\ee
The limit $r_H=0$ then corresponds either to the case of EMFLS boson stars, where $Q=gQ_N$, or to the limit of extremal black holes, since we work with isotropic coordinates. 
To obtain a quantitative measure of the scalar hair of the black holes, we make use of the hairiness parameter $h$ \cite{Herdeiro:2020xmb}
\be
h= 1- \frac{Q_H}{Q}= \frac{g\, Q_N}{Q} \, .
\ee
For boson stars $h=1$, whereas for RN black holes $h=0$, while charged hairy black holes possess hairiness values in between these limits.

Further physically interesting characteristics of the event horizon are the Hawking temperature $T_H$, which is proportional to the surface gravity $\kappa^2=-\frac12 \nabla_\mu \xi_\nu \nabla^\mu \xi^\nu $ and the horizon area $A_H$, related to the entropy of the black hole as
$S=\frac{\pi A_H}{\alpha^2}$
\be
T_H=\frac{\kappa}{2\pi}= \frac{1}{16 \pi r_H}e^{f_0-f_1}\biggl.\biggr|_{r=r_H}\, , \quad
A_H=64 \pi r_H^2 e^{f_0+f_1}\biggl.\biggr|_{r=r_H} \, .
\label{temparea}
\ee
The value of the electrostatic potential on the event horizon $A_0(r_H)=\omega/g$ defines the chemical potential $\mu_{ch}$ of the solutions, $\mu_{ch}=A_0(r_H)$.

\section{Numerical results}

The set of five coupled ordinary differential equations for the functions $X,Y,A_0,F_0, F_1$, obtained after substitution of the Ansatz \re{metric},\re{scalans},\re{Aans} into the general system of equations \re{field-eqs-em},\re{field-eqs-scalar}, has been solved numerically subject to the boundary conditions discussed above.
We have made use of a sixth-order finite difference scheme, where the system of equations is discretized on a grid with a typical size of 529 points in radial direction. 
To facilitate the calculations in the near horizon area, we have made use of the reparametrization of the metric functions \re{metric-hor} and introduced the new compact radial coordinate $x=\frac{r-r_H}{c+r}$, where $c$ is an arbitrary constant used to adjust the contraction of the grid.
The emerging system of nonlinear algebraic equations has been solved using the Newton-Raphson scheme. 
Calculations have been performed with the packages FIDISOL/CADSOL \cite{schoen}, with typical errors of order of $10^{-4}$.

With our choice of parametrization and gauge fixing condition, the input parameters are the scalar masses $m$ and $\mu$, the gravitational coupling $\alpha$, the gauge coupling $g$, the horizon radius $r_H$, and the angular frequency $\omega$. 
For the sake of compactness we now fix $m=1$ and, following our previous investigation of boson stars in the EMFLS model \cite{Kunz:2021mbm}, we consider two physically distinct cases, $\mu^2 = 0.25$ and $\mu=0$. 
Further, we fix the value of the gauge coupling $g=0.1$, then the variation of the angular frequency is directly related to the change of the chemical potential $\mu_{ch}$.

\subsection{Gauged EMFLS Boson Stars}

We  start by recalling the properties of regular spherically symmetric
gauged boson stars in the EMFLS model \cite{Kunz:2021mbm}.
These gauged EMFLS boson stars exist within a restricted interval of values of the angular frequency $\omega \in [\omega_{min},~\omega_{max}]$. 
The maximal value $\omega_{max}$ corresponds to the mass of the excitations of the complex scalar component $\phi$, i.e., our choice of parameters yields $\omega_{max}=1$. 
As the angular frequency is decreased below this maximal value, a first branch of the gauged boson stars arises from the electrostatic vacuum.  
Along this branch the properties of the gauged boson stars are not very different from the corresponding solutions in the ungauged limit, and the electromagnetic energy remains much smaller than the total energy of the configuration.  
As $\omega$ decreases, this branch extends backward up to a minimal value of the frequency $\omega_{min}$ where a bifurcation occurs and a second, forward branch is encountered.
The subsequent pattern depends on the relative strength of the competing gravitational and electrostatic interactions.
The $U(1)$ gauged boson stars then exhibit either the typical spiraling and oscillating pattern of ungauged boson stars, or their behavior resembles the properties of the gauged Q-balls in flat spacetime \cite{Loiko:2019gwk}.
In the latter case, the complex charged component $\phi$ becomes massless inside the core of the configuration, the electrostatic long-range repulsive force inflates the interior region, and the size of the gauged boson star rapidly increases. 
In the former case, the long-range gravitational attraction becomes stronger than the electrostatic force, the complex component $\phi$ remains massive in the interior region, and the gauged boson star exhibits an inspiraling scenario towards a limiting solution.     

This picture changes dramatically as the mass of the real scalar component vanishes such that it becomes long-ranged. 
The presence of two long-range fields violates the delicate force balance between the gravitational attraction, the electrostatic repulsion and the scalar interactions.
As a result there is only a single branch of regular gauged EMFLS solutions, which terminates at a minimal value of the angular frequency $\omega_{min}$ \cite{Kunz:2021mbm}. 
The mass and the electric charge of the EMFLS boson stars increase monotonically along this branch, and both diverge as the minimal frequency is approached. 
The maximal frequency corresponds to the mass threshold of the complex field, in such a limit the fields approach the vacuum although the size of the Q-ball rapidly increases. 

\subsection{Reissner-Nordstr\"{o}m  black holes with resonant two-component scalar hair: Finite mass parameter $\mu^2 \neq 0$}

\begin{figure}[h!]
\begin{center}
\includegraphics[height=.33\textheight,  angle =-90]{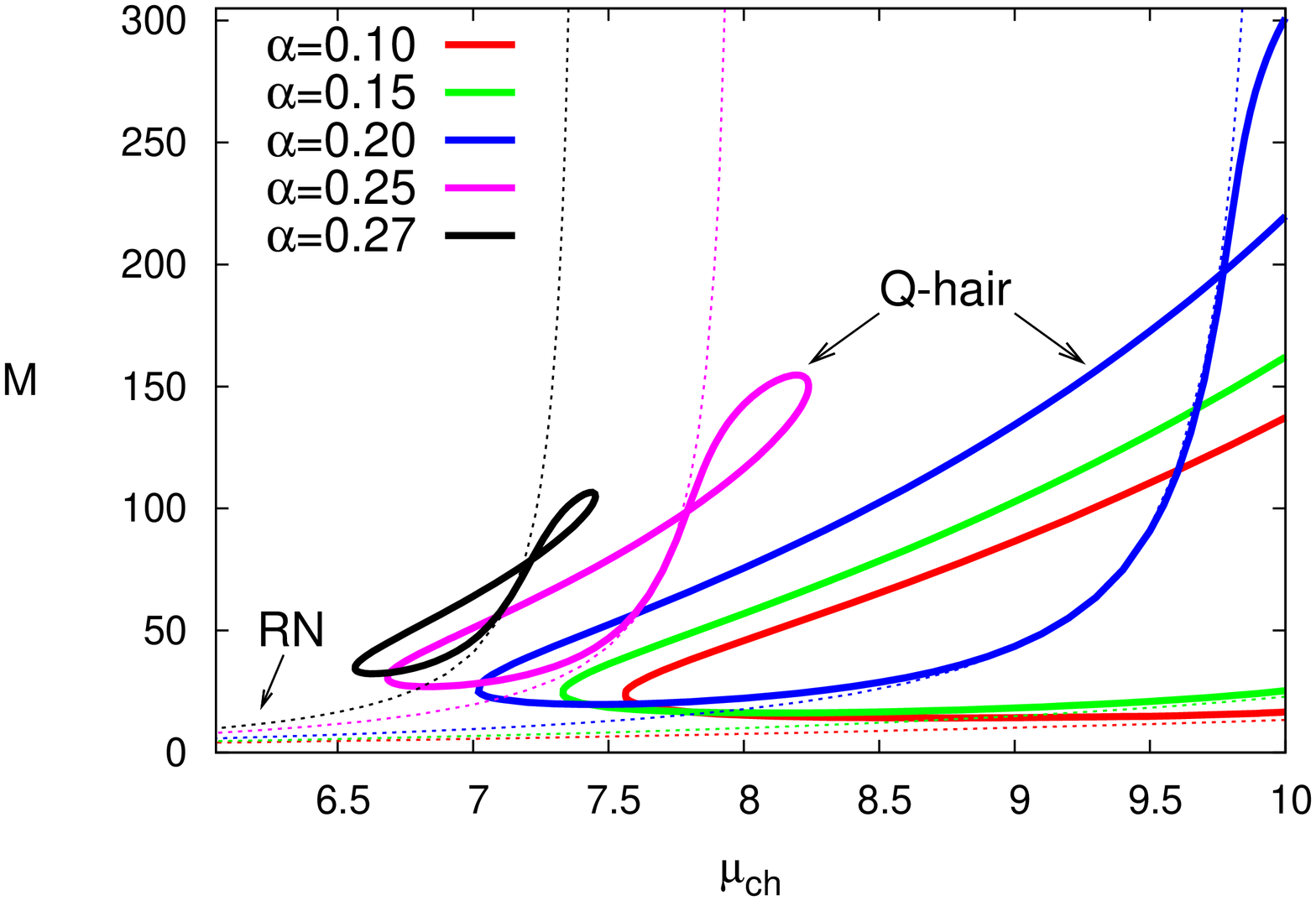}
\includegraphics[height=.33\textheight,  angle =-90]{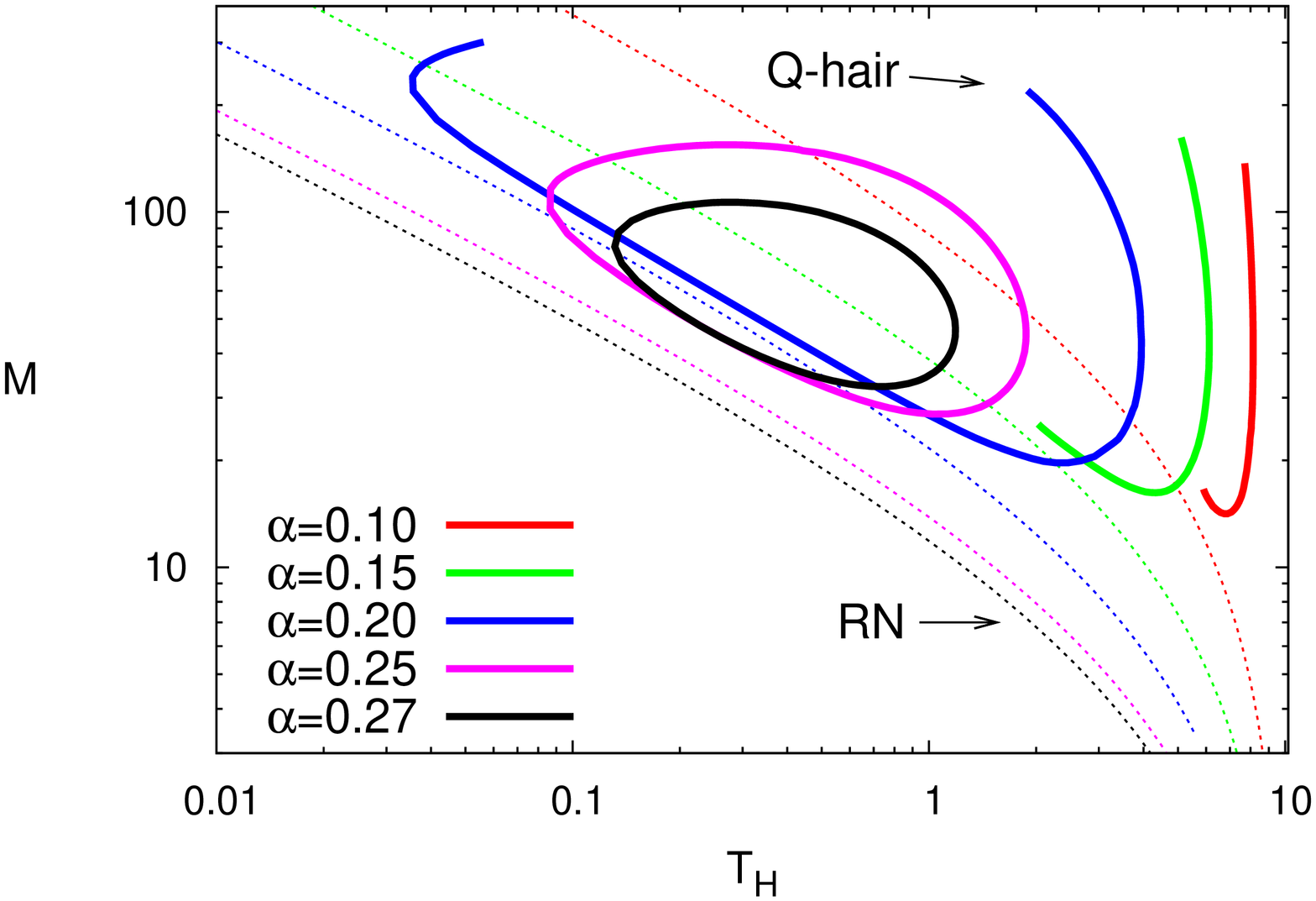}
\includegraphics[height=.33\textheight,  angle =-90]{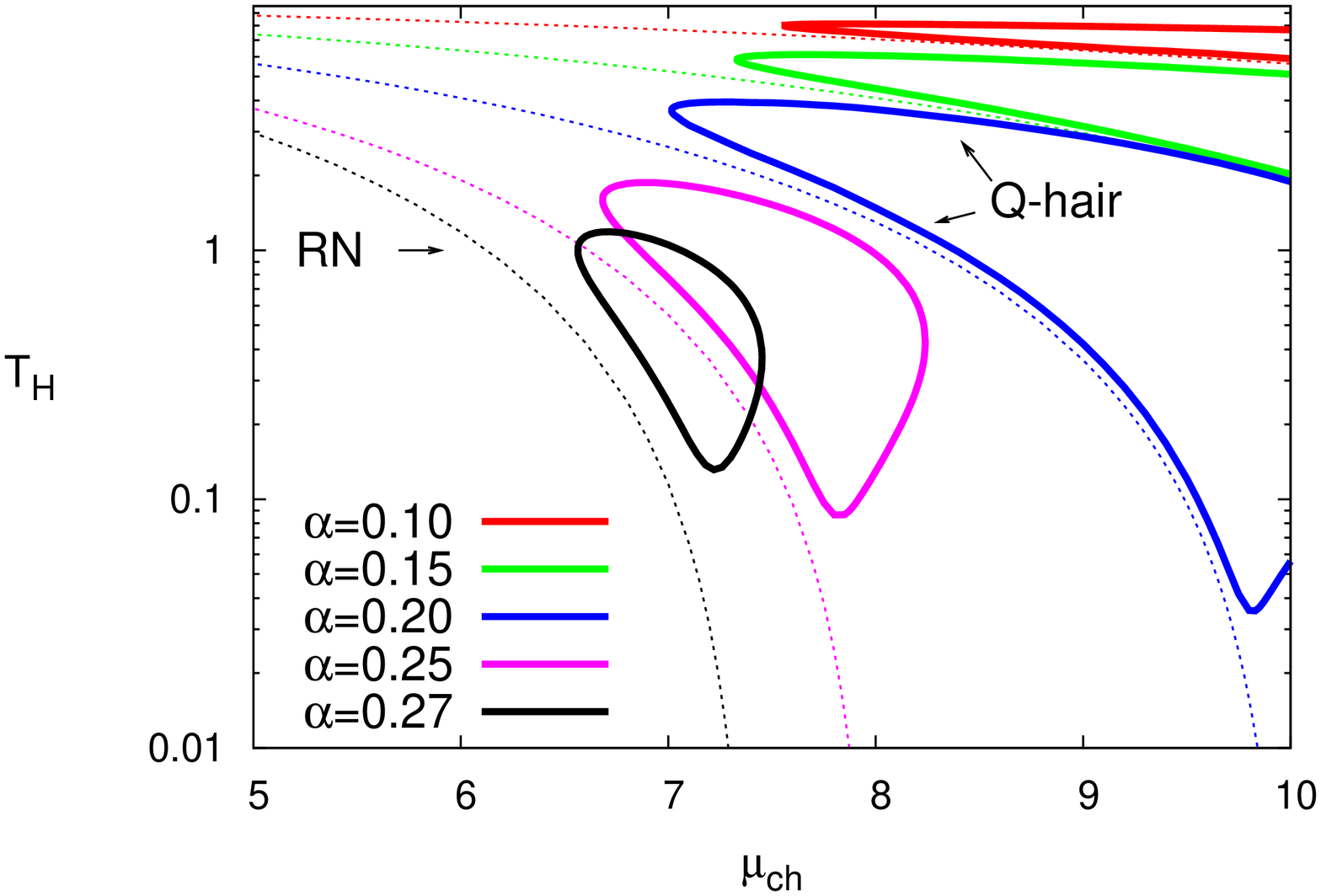}
\includegraphics[height=.33\textheight,  angle =-90]{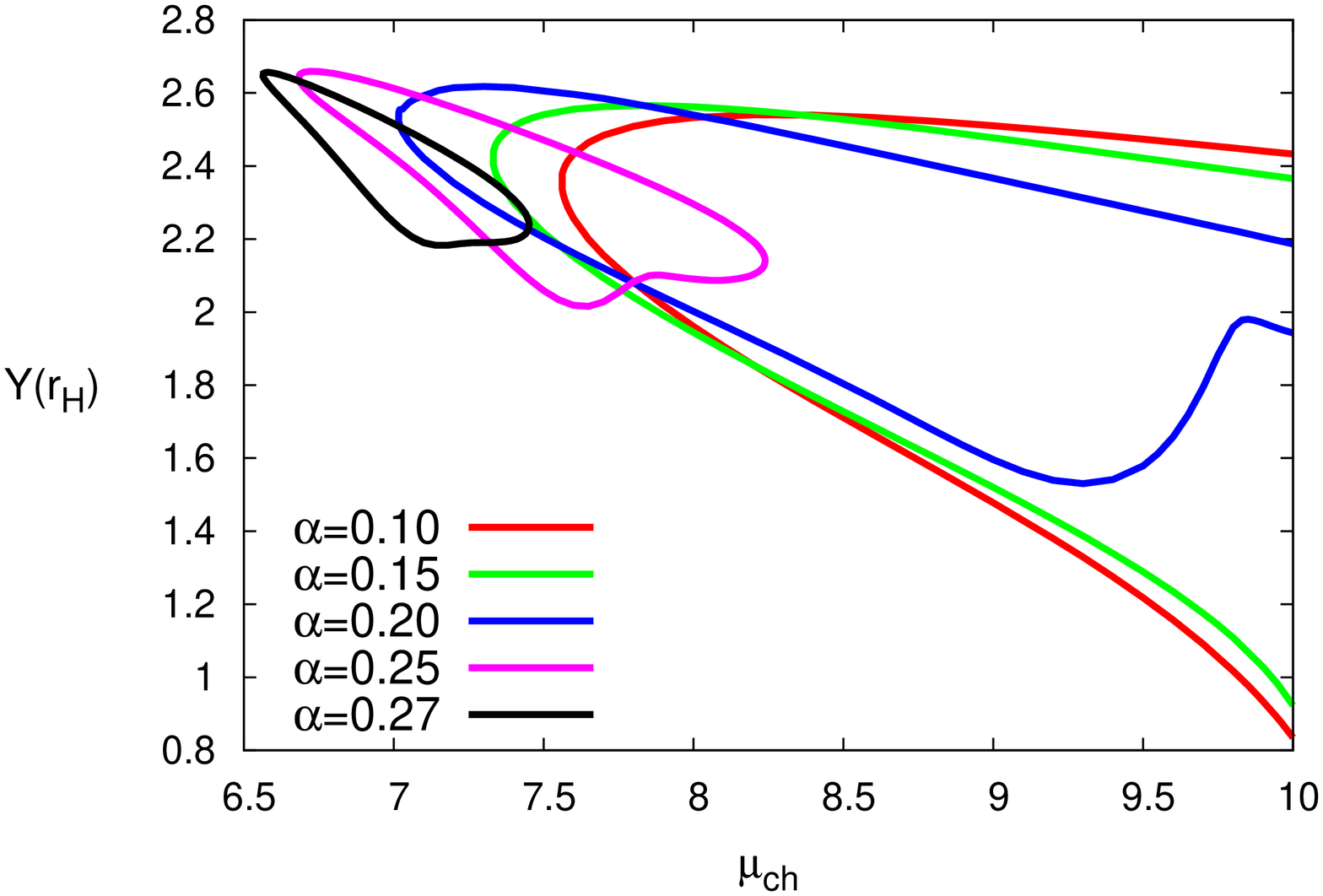}
\includegraphics[height=.33\textheight,  angle =-90]{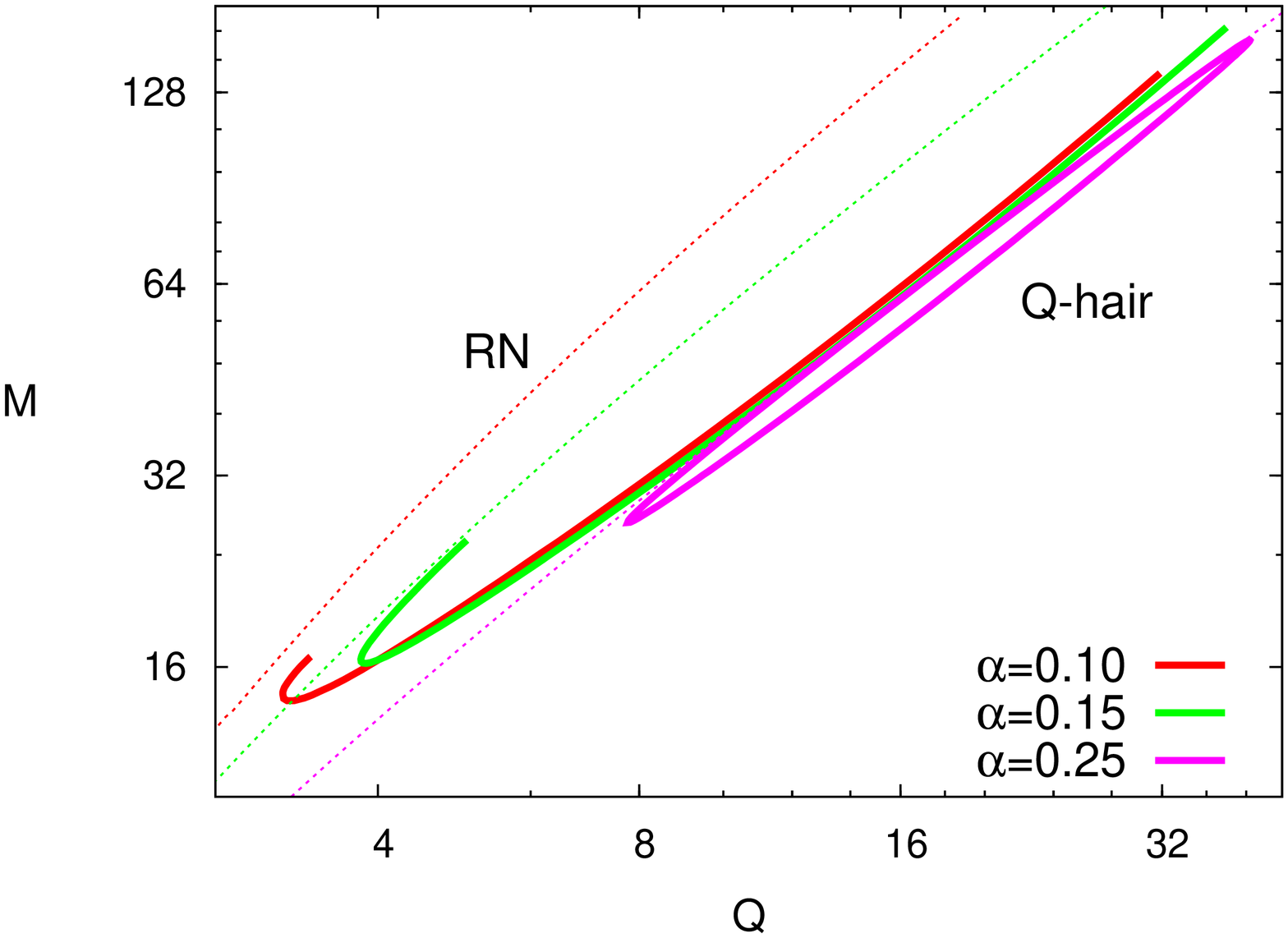}
\includegraphics[height=.33\textheight,  angle =-90]{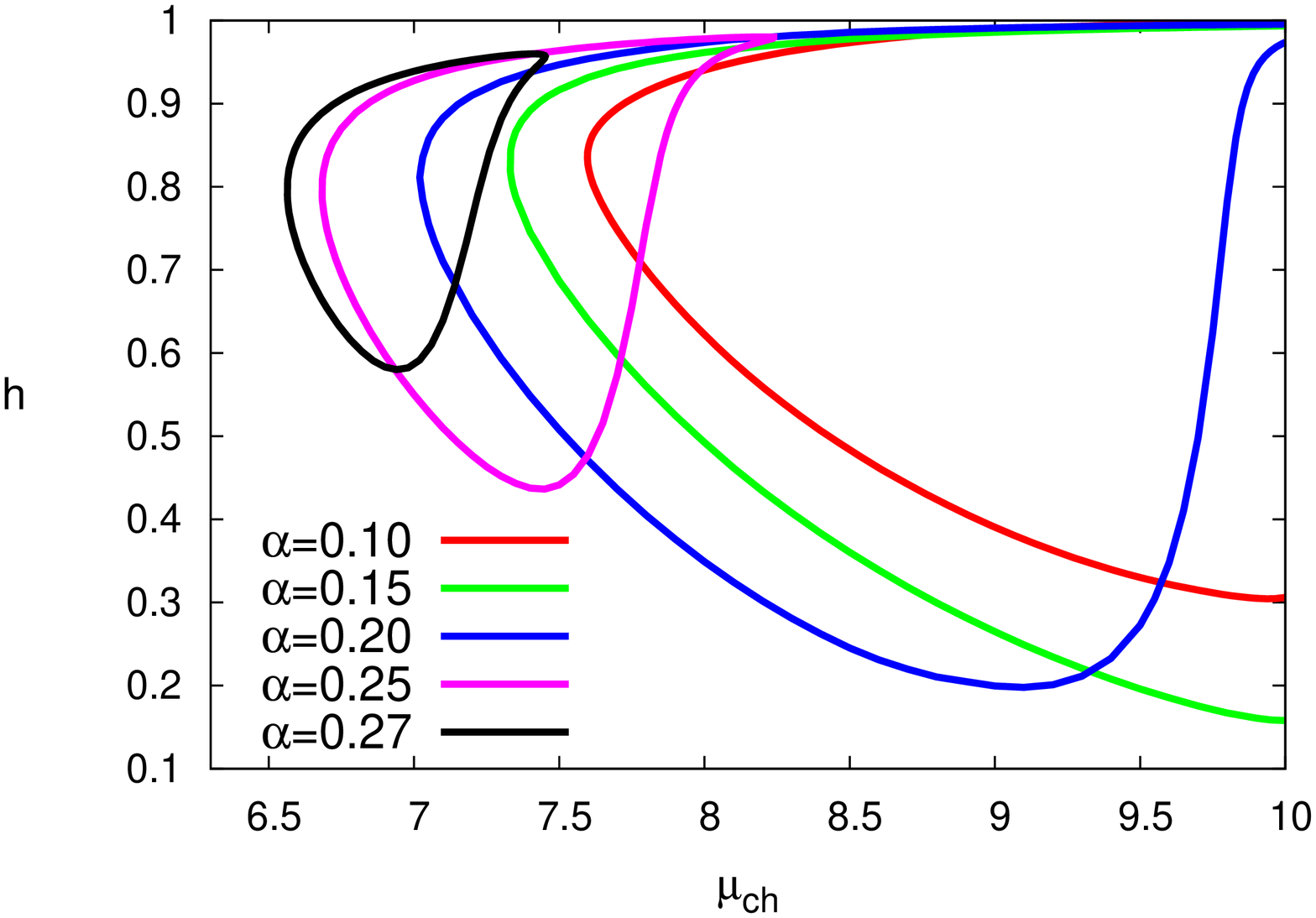}
\end{center}
\caption{\small
Charged EMFLS black holes with resonant Q-hair: 
Mass $M$ vs chemical potential $\mu_{ch}$ (upper left) and vs Hawking temperature $T_H$ (upper right); Hawking temperature $T_H$ (middle left) and horizon value of profile function $Y(r_H)$ (middle right) vs chemical potential $\mu_{ch}$; and mass $M$ vs charge $Q$ (lower left, note the log base 4) and hairiness $h$ vs chemical potential $\mu_{ch}$ (lower right) for a set of values of the gravitational coupling $\alpha$, for horizon radius $r_H=0.1$, gauge coupling $g=0.1$ and mass parameter $\mu^2=0.25$. 
For comparison the corresponding RN properties are shown.
}
    \lbfig{fig1}
\end{figure}

In analogy to charged black holes with resonant single-component scalar Q-hair \cite{Herdeiro:2020xmb,Hong:2020miv}, a small event horizon can be immersed at the center of the gauged EMFLS boson stars. 
Since the event horizon is endowed with the electric charge $Q_H$
an electrostatic repulsive force appears acting on the charged $Q$-cloud surrounding the black hole.
This clearly affects the pattern described above as seen in the following. 

We exhibit in Fig.~\ref{fig1} (upper left) the total mass $M$ of the charged hairy black holes versus their chemical potential $\mu_{ch}$ for a set of values of the gravitational coupling $\alpha$ and the fixed isotropic horizon radius $r_H=0.1$, the gauge coupling $g=0.1$ and the mass parameter $\mu^2=0.25$. 
For relatively weak gravitational coupling $\alpha$, the presence of the event horizon does not seriously affect the pattern outlined above for the regular solutions. 
Along the fundamental branch the mass of the EMFLS black holes with resonant scalar hair is slightly higher than the mass of the corresponding RN black holes, while getting closer to it with increasing $\mu_{ch}$. 
On the upper (electrostatic) branch the mass of the Q-hairy black holes is much higher, and the corresponding hairiness $h$ is significantly larger than on the lower (scalar) branch, as shown in Fig.~\ref{fig1} (lower right).

The overall pattern changes as the gravitational coupling increases. 
Gravity then prevents an inflationary expansion of the Q-clouds with increasing mass, so the upper critical value of the angular frequency is shifted down from the mass threshold. 
On the other hand, the hairy EMFLS black holes cannot enter the spiralling pattern because of the electrostatic repulsive force acting on the charged Q-clouds from the horizon of the black holes with horizon charge $Q_H$ \re{Q_Q_H}. 
This changes the lower bound on the possible range of values of the chemical potential. 
Thus, there is a particular value of the chemical potential which corresponds to a balance of the competing forces of gravitational attraction and electrostatic repulsion. 
Consequently, the mass of the hairy EMFLS black holes coincides with the mass of the 
RN black holes at a particular value of the chemical potential that decreases with increasing $\alpha$, as seen in Fig.~\ref{fig1} (upper left). 
Since the allowed range of values of the chemical potential decreases as the gravitational coupling increases, the hairy EMFLS black holes cease to exist at some critical value of $\alpha$. 

The Hawking temperature $T_H$ of the hairy EMFLS black holes is always higher that of the corresponding RN black holes, whose domain of existence is much larger (see Fig.~\ref{fig1} (middle left)). 
In this figure we also display the dependence of the mass on the Hawking temperature (upper right) and the values of the horizon value of the scalar profile function $Y(r_H)$ versus the chemical potential $\mu_{ch}$ (middle right). 
In addition, the figure shows the mass $M$ versus the charge $Q$ (lower left) and the hairiness $h$ versus the chemical potential $\mu_{ch}$ (lower right).
The latter diagram clearly shows, that these black holes with resonant Q-hair never reach the RN limit with $h=0$.

\begin{figure}[h!]
\begin{center}
\includegraphics[height=.33\textheight,  angle =-90]{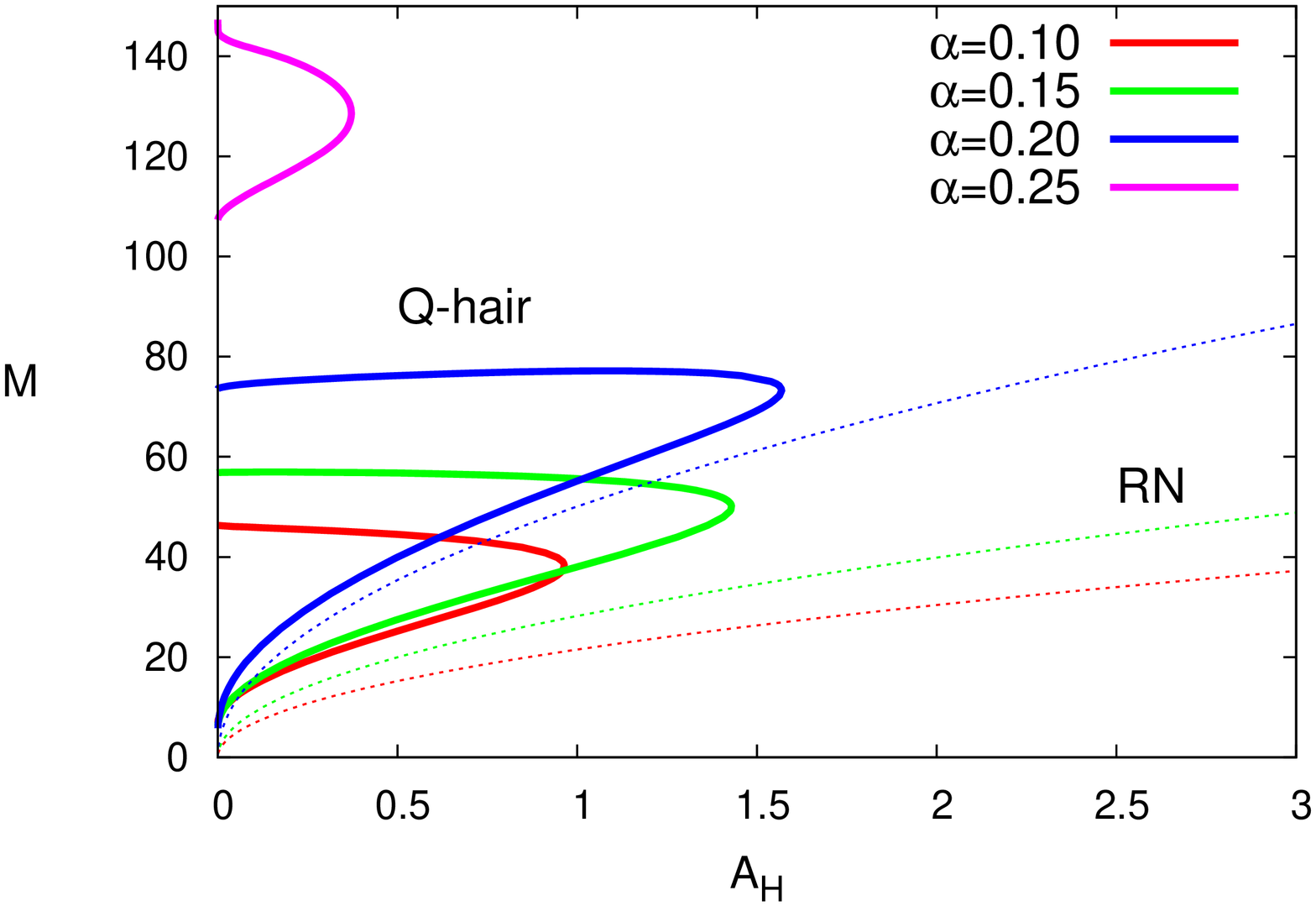}
\includegraphics[height=.33\textheight,  angle =-90]{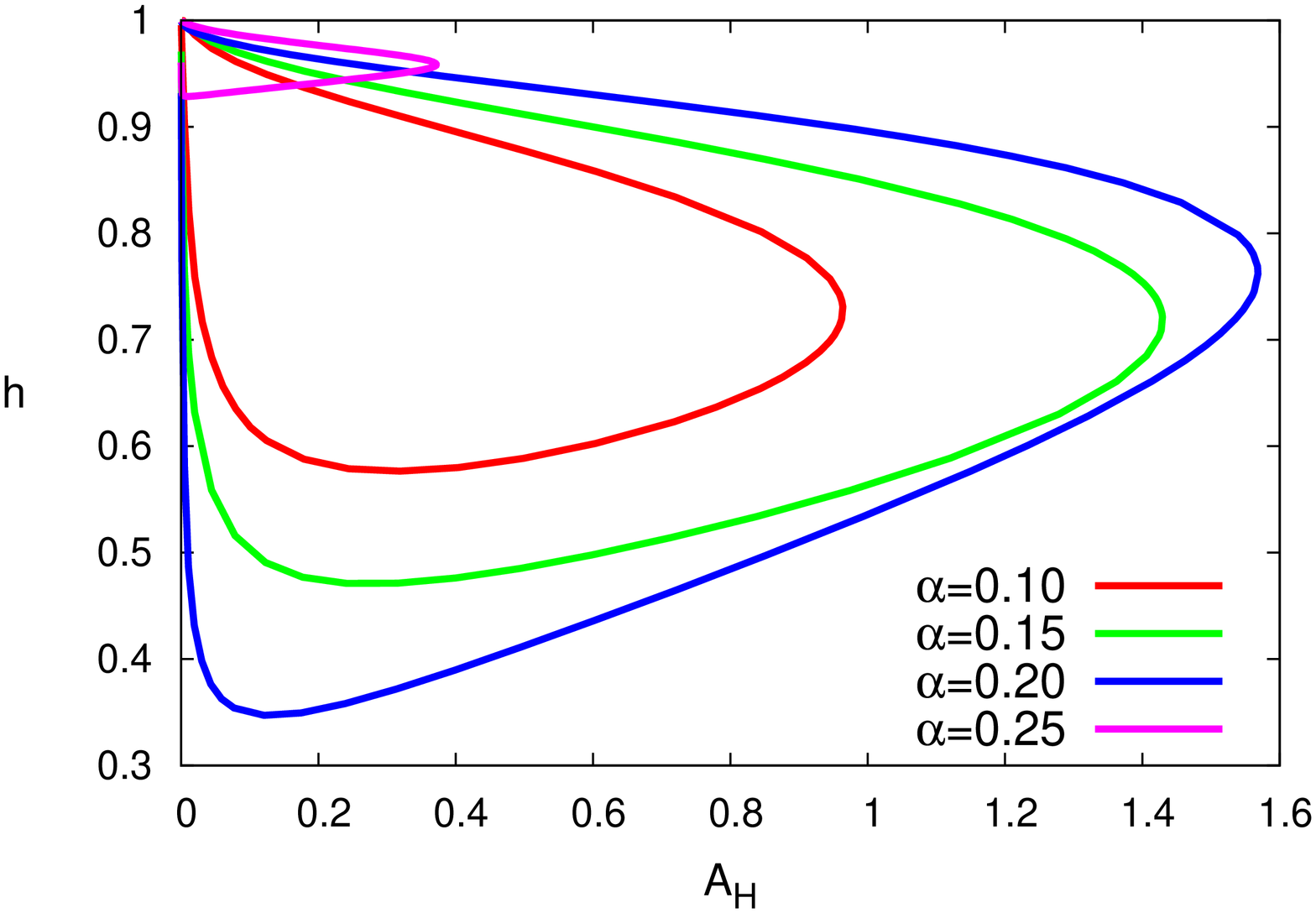}
\end{center}
\caption{\small
Charged EMFLS black holes with resonant Q-hair:
Mass $M$ (left) and hairiness (right) vs horizon area $A_H$ for  fixed chemical potential $\mu_{ch}=8$ and a set of values of the gravitational coupling $\alpha$, for gauge coupling $g=0.1$ and mass parameter $\mu^2=0.25$. 
}
   \lbfig{fig2}
\end{figure}

Figure \ref{fig2} exhibits the ADM mass of the hairy EMFLS black holes and the hairiness parameter $h=Q_N/Q$ in a grand canonical ensemble, i.e., for a fixed chemical potential, $\mu_{ch}=8$, versus the horizon area $A_H$ for a set of values of the gravitational coupling $\alpha$ and for gauge coupling $g=0.1$ and mass parameter $\mu^2=0.25$. 
The Q-hairy black holes emerge smoothly from the lower branch of the corresponding EMFLS boson stars \cite{Kunz:2021mbm}.
The fundamental branch of Q-hairy black holes then exists up to a maximal value of the horizon area $A_H^{max}$. 
The mass $M$ and the charge $Q$ of the configuration increase along this branch. 
At the maximal value $A_H^{max}$ the fundamental branch bifurcates with the secondary branch leading back to a limiting solution.
This is either the limiting boson star solution on the electrostatic branch, or it corresponds to an apparently singular solution.
The latter might be related to the intriguing solution, where the spacetime splits into two parts, with the inner part confining the scalar fields in a very small region and the outer part representing an extremal RN black hole (see the discussion in \cite{Brihaye:2020vce,Brihaye:2021phs,Brihaye:2021mqk}).

The hairiness parameter $h$ gradually decreases along the lower (mass) branch.
Along the secondary branch, it continues to decrease at first.
If the end point is a boson star, it then reaches a minimal value at some critical value of $A_H$, from where it rapidly grows and approaches the boson star limit $h=1$ again, as seen in Fig.~\ref{fig2} (right). 
Otherwise, it continues to decrease as it approaches the hairiness value of the limiting solution. 
Unfortunately, our use of isotropic coordinates precludes a clear analysis of the limiting behavior.
Similar to the hairy electrostatic black holes in a single-component model \cite{Herdeiro:2020xmb,Hong:2020miv}, there is always a gap between the hairy black holes and the RN black holes for a finite value of the mass parameter $\mu$.

\begin{figure}[h!]
\begin{center}
\includegraphics[height=.33\textheight,  angle =-90]{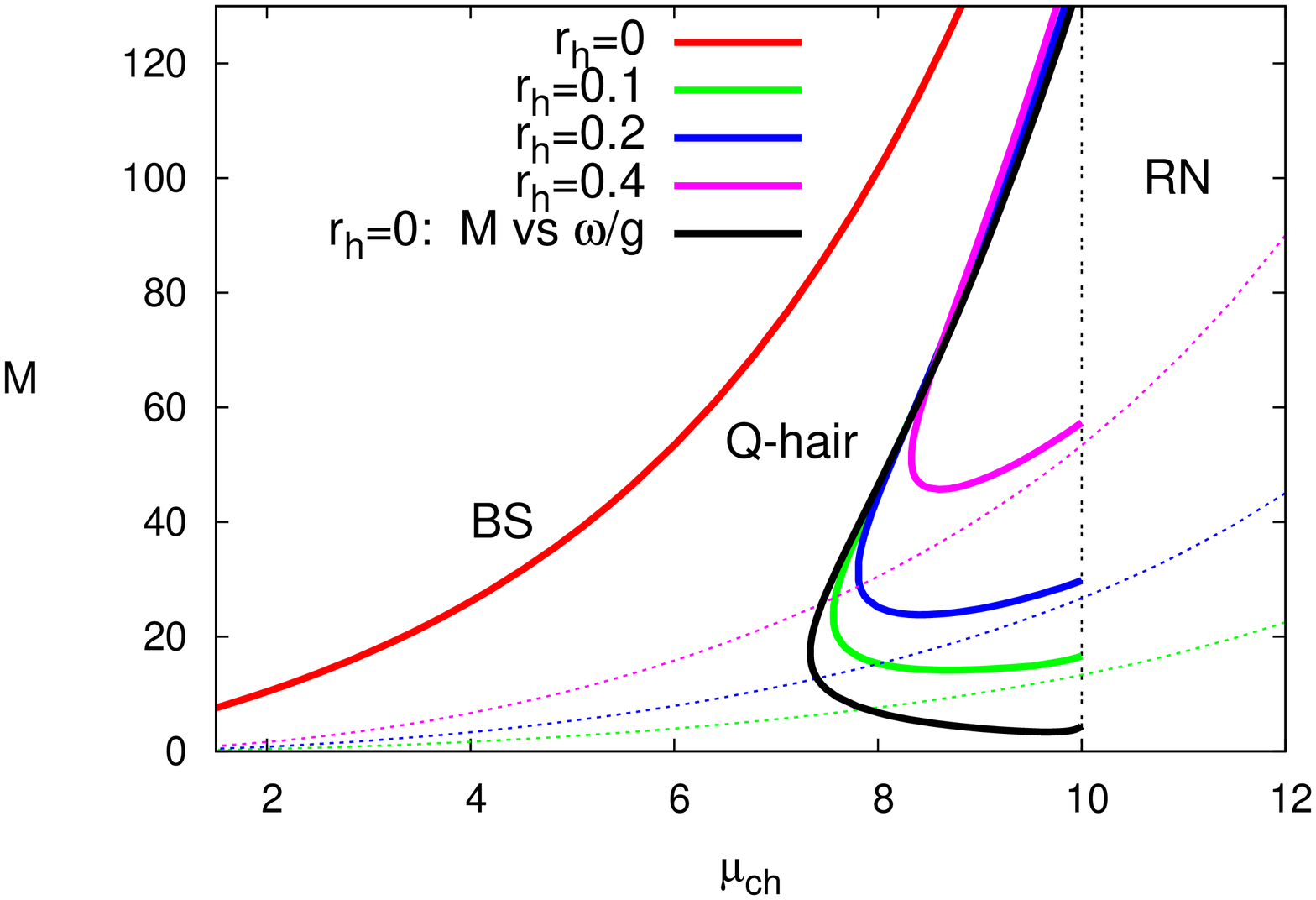}
\includegraphics[height=.33\textheight,  angle =-90]{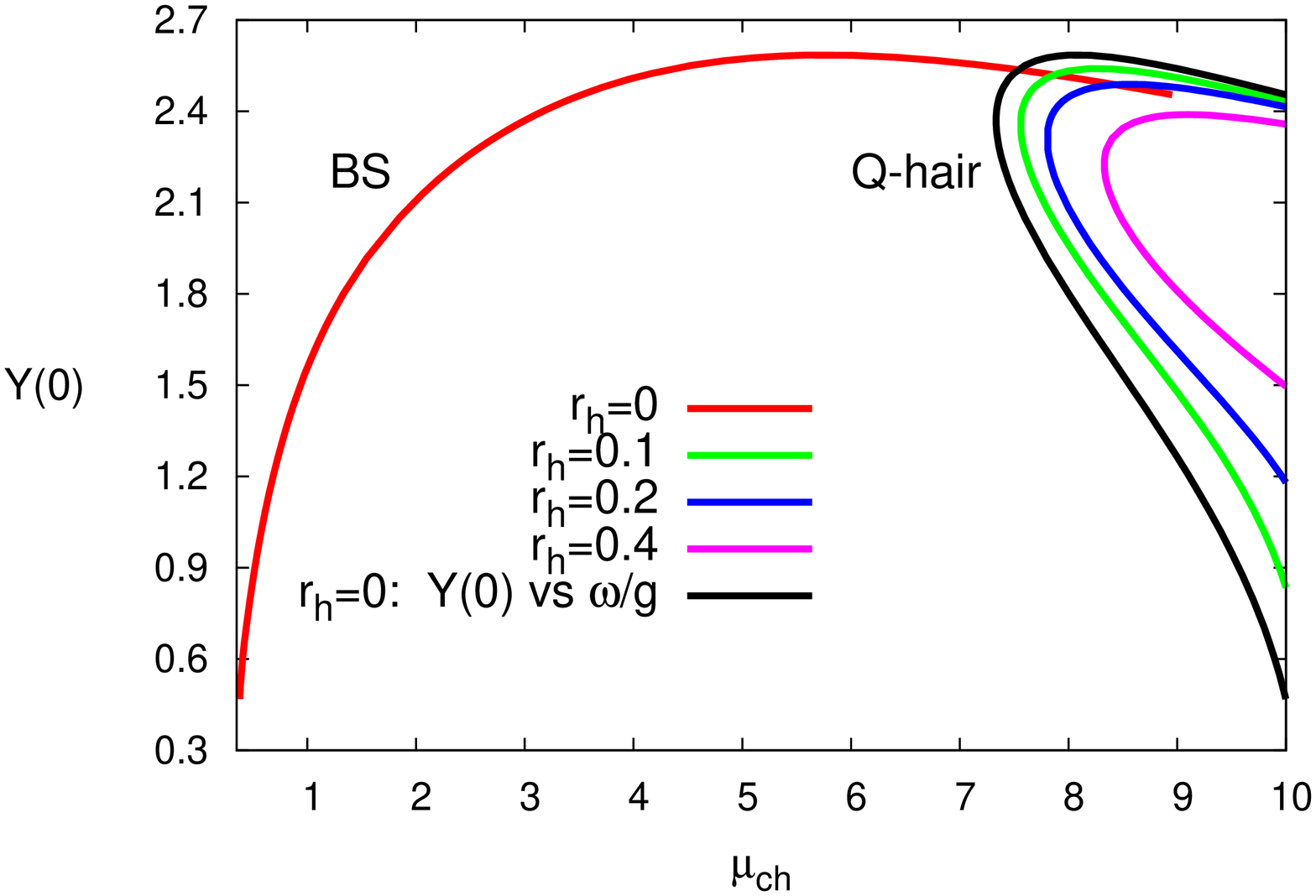}
\end{center}
\caption{\small
Phase structure of EMFLS black holes: Mass $M$ (left) and horizon value of scalar function $Y(r_H)$ (right) vs chemical potential $\mu_{ch}$ for a set of values of the horizon radius $r_H$, for gravitational coupling $\alpha=0.1$, gauge coupling $g=0.1$ and mass parameter $\mu^2=0.25$.
$M$ and $Y(r_H)$ are also shown for boson stars versus $\mu_{ch}$ (red) and versus scaled frequency $\omega/g$ (black).
For comparison the corresponding RN properties are shown (left).
}
    \lbfig{fig3}
\end{figure}

Considering the dependence of the hairy EMFLS black holes on the horizon radius $r_H$ we observe that, for a given value of the gravitational coupling $\alpha$, the allowed range of values of the chemical potential decreases as $r_H$ increases, as seen in Fig.~\ref{fig3} (left), where we have exchanged the roles of $\alpha$ and $r_H$ with respect to Fig.~\ref{fig1}.
Indeed, the increase of the horizon radius yields a significant ($\sim r_H^2$) growth of the electric horizon charge $Q_H$. 
Thus the electric repulsive force acting on the Q-cloud becomes stronger, and for a certain critical value of $r_H$ the hairy EMFLS  black holes cease to exist. 
Figure \ref{fig3} (right) demonstrates the horizon value of scalar function $Y(r_H)$ versus the chemical potential $\mu_{ch}$ for the same set of solutions.

The boson star limit ($r_H=0$) is also exhibited in Fig.~\ref{fig3}, employing two different ways.
The red curves employ the chemical potential, i.e., the value of the gauge potential at the origin, thus demonstrating the discontinuity of this quantity in the limit $r_H=0$.
The black curves, on the other hand, employ the angular frequency of the boson stars divided by the gauge coupling, thus mimicking the resonance condition of the Q-hairy black holes.
Here we see a smooth convergence of the properties in the limit $r_H=0$.

\subsection{Reissner-Nordstr\"{o}m  black holes with resonant two-component scalar hair:  massless limit $\mu = 0$}

\begin{figure}[p!]
\begin{center}
\includegraphics[height=.33\textheight,  angle =-90]{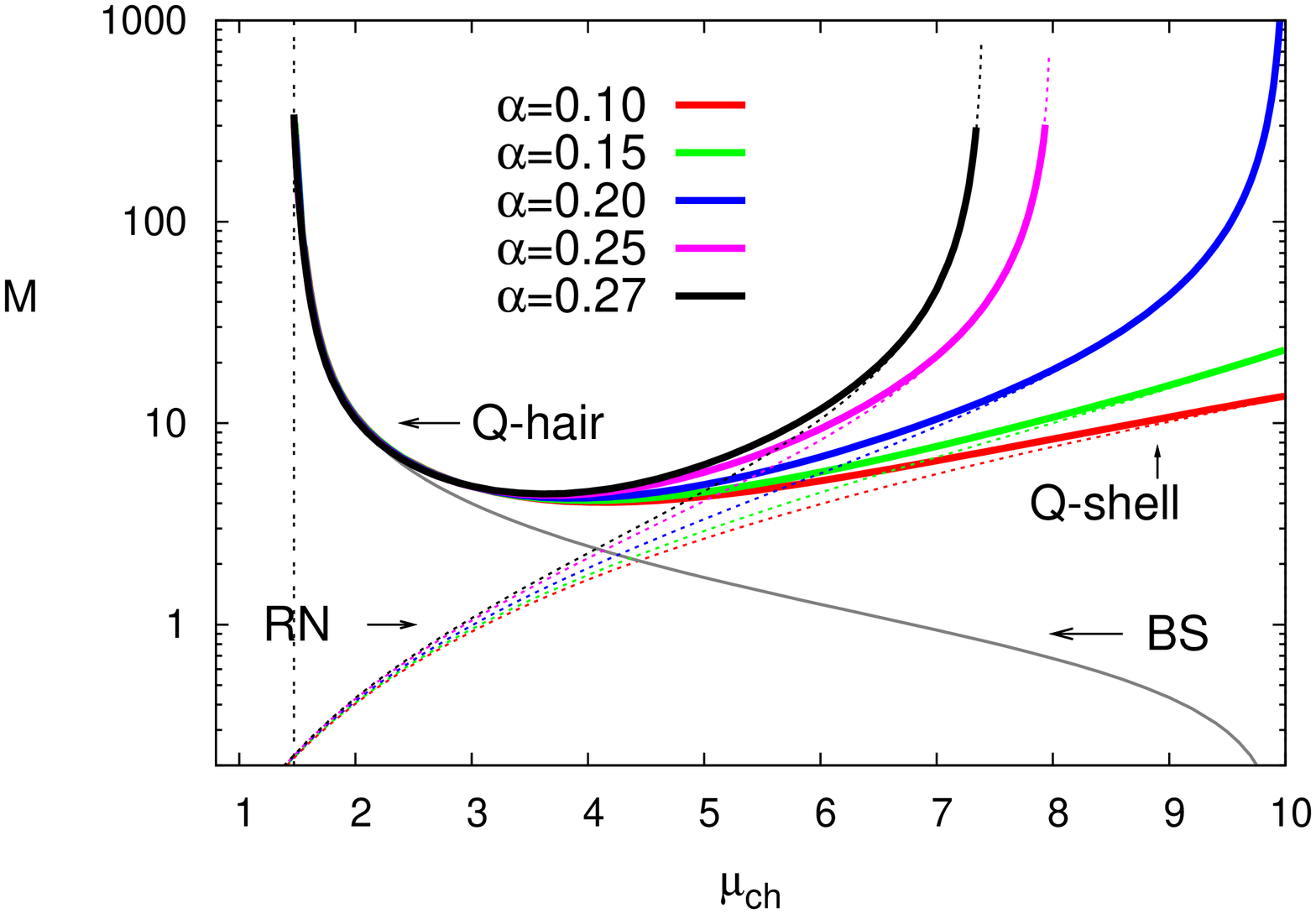}
\includegraphics[height=.33\textheight,  angle =-90]{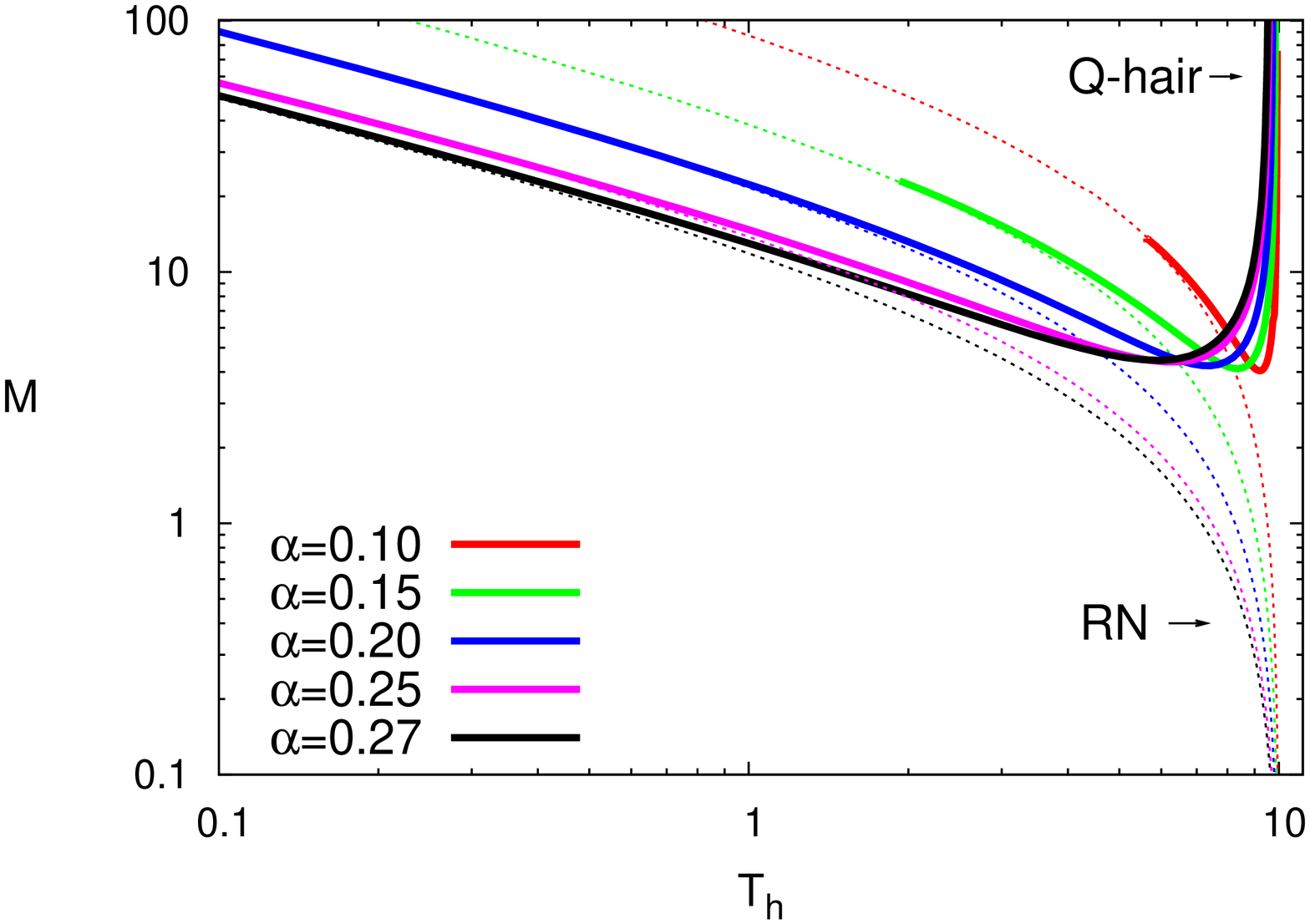}
\includegraphics[height=.33\textheight,  angle =-90]{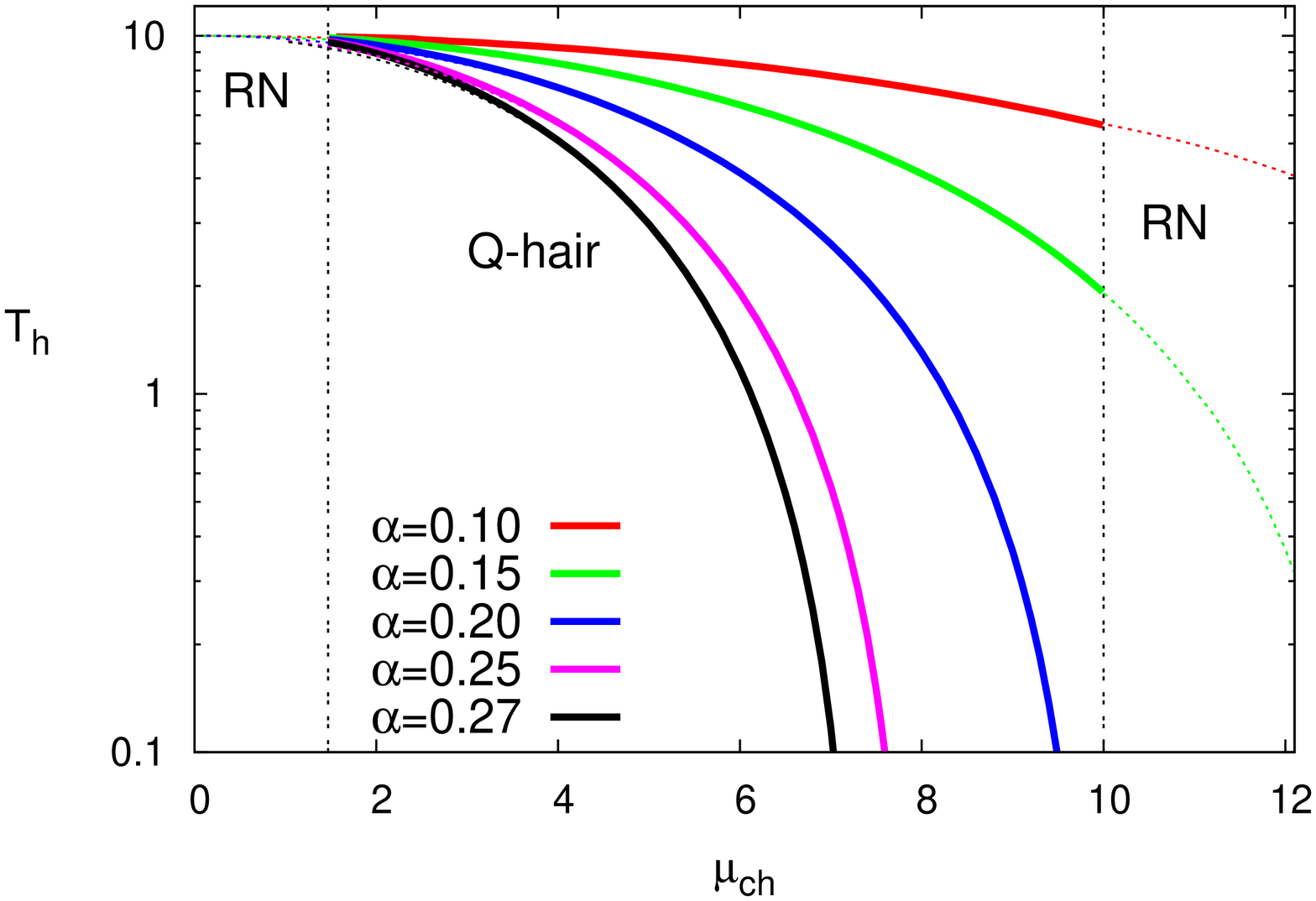}
\includegraphics[height=.33\textheight,  angle =-90]{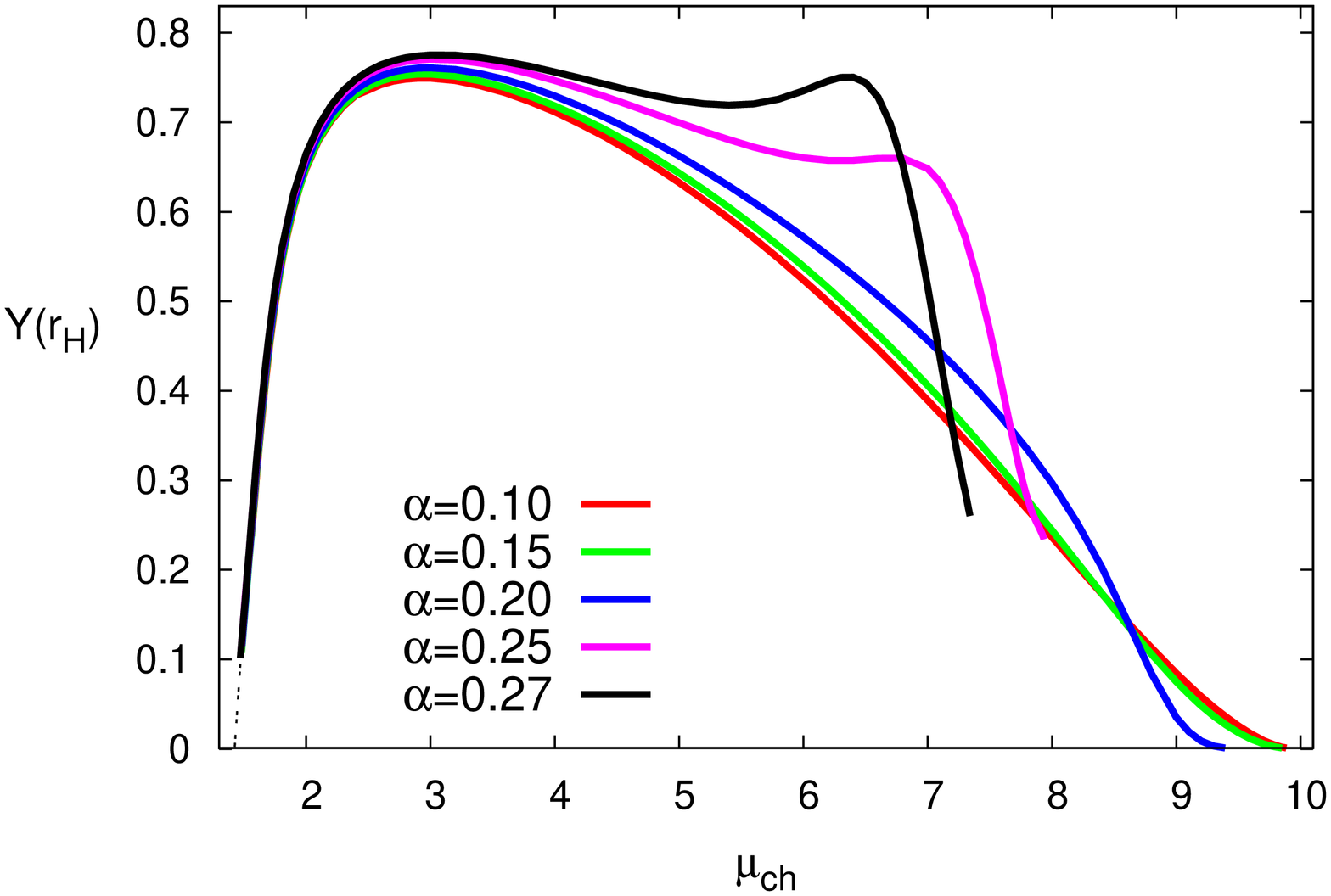}
\includegraphics[height=.33\textheight,  angle =-90]{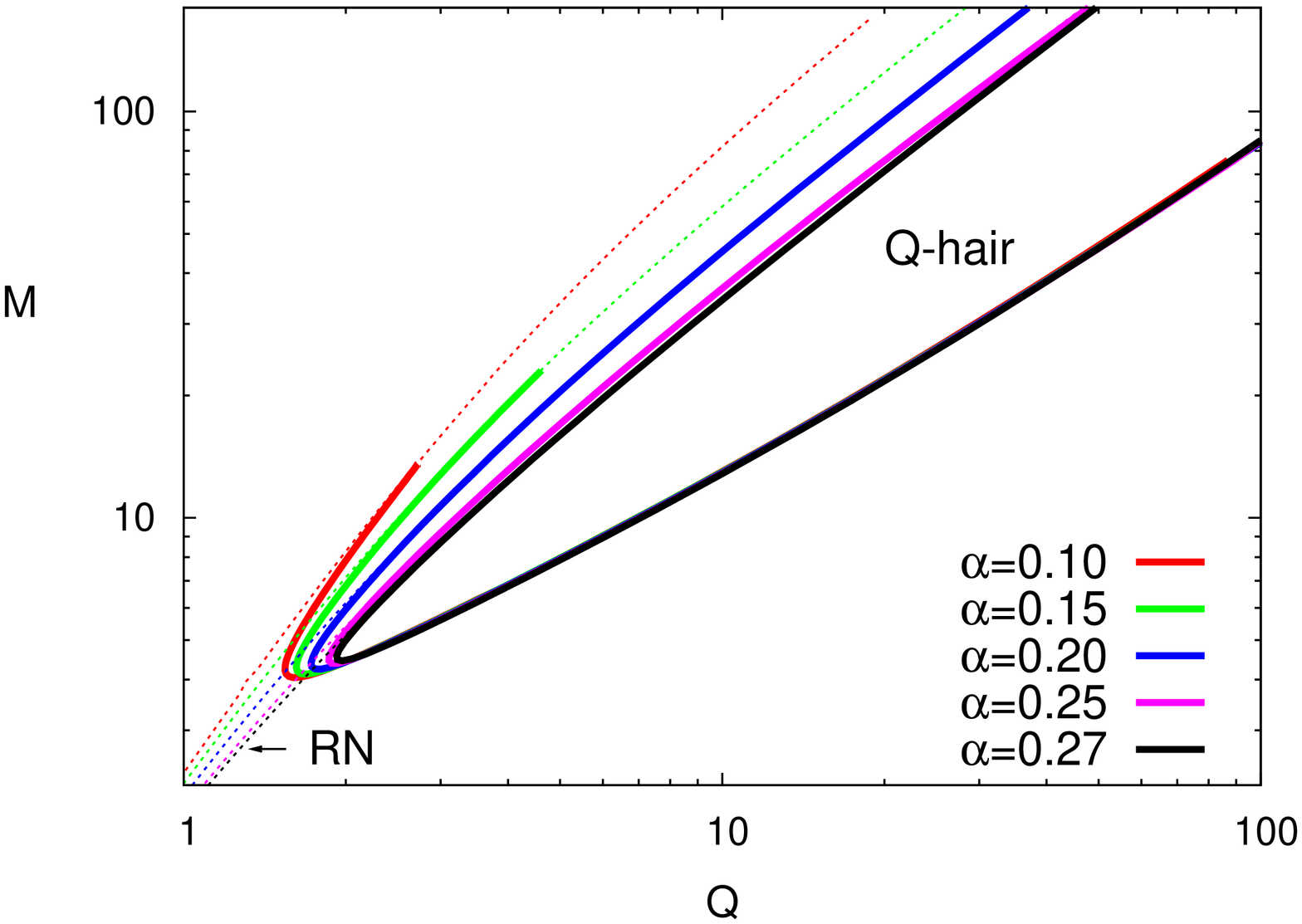}
\includegraphics[height=.33\textheight,  angle =-90]{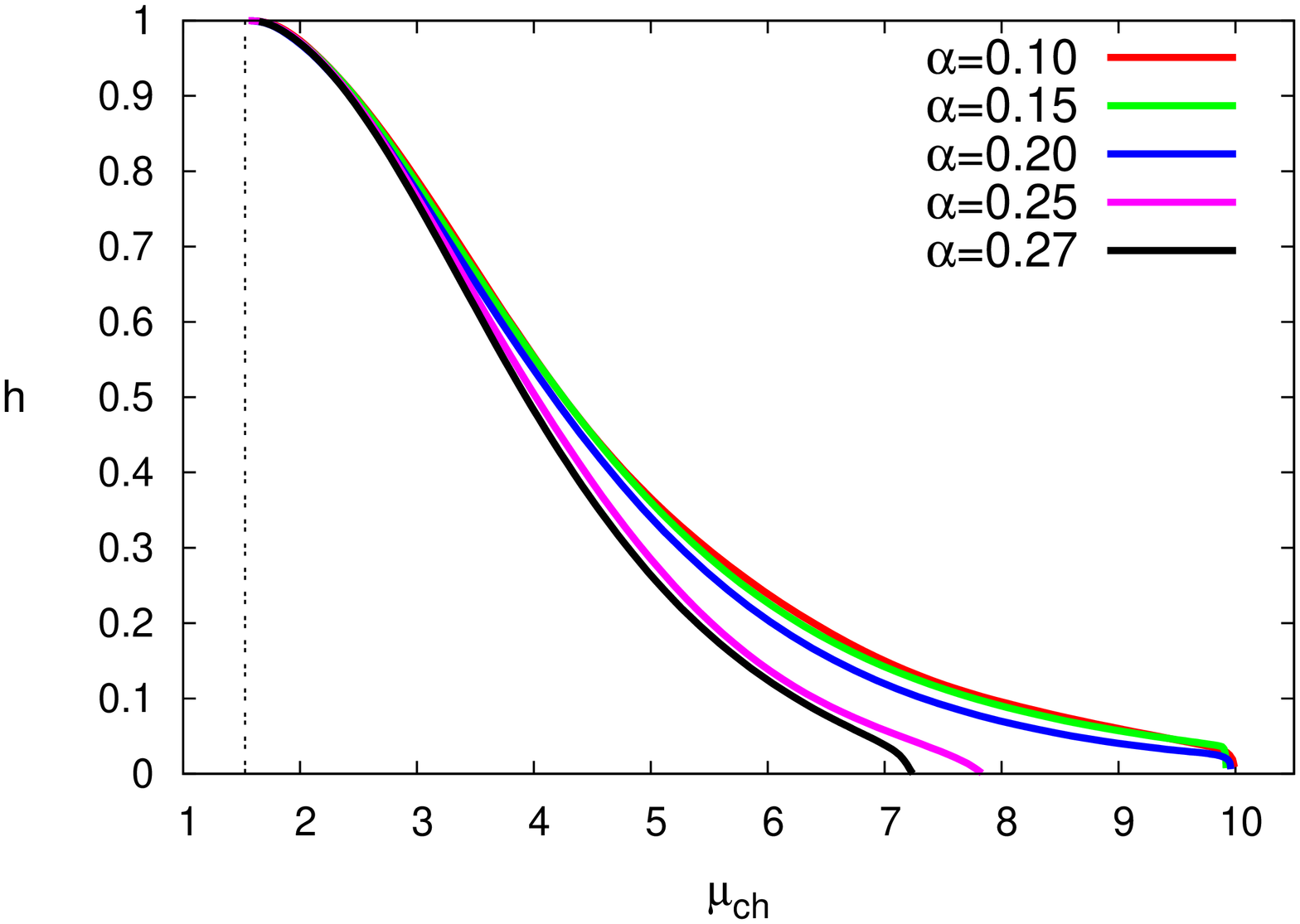}
\end{center}
\caption{\small
Charged EMFLS black holes with resonant Q-hair in the massless ($\mu=0$) limit:
Mass $M$ vs chemical potential $\mu_{ch}$ (upper left) and vs Hawking temperature $T_H$ (upper right); Hawking temperature $T_H$ (middle left) and horizon value of profile function $Y(r_H)$ (middle right) vs chemical potential $\mu_{ch}$; mass $M$ vs charge $Q$ (lower left), and hairiness $h$ vs chemical potential $\mu_{ch}$ (lower right) for a set of values of the gravitational coupling $\alpha$, for horizon radius $r_H=0.1$, gauge coupling $g=0.1$. 
For comparison the corresponding RN properties are shown.
For boson stars $M$ is also shown versus scaled frequency $\omega/g$ (grey, upper left).
The vertical lines represent the limits of the domain of existence ($\mu_{min}\approx 1.47$).
}
    \lbfig{fig4}
\end{figure}

\begin{figure}[h!]
\begin{center}
\includegraphics[height=.33\textheight, angle =-90]{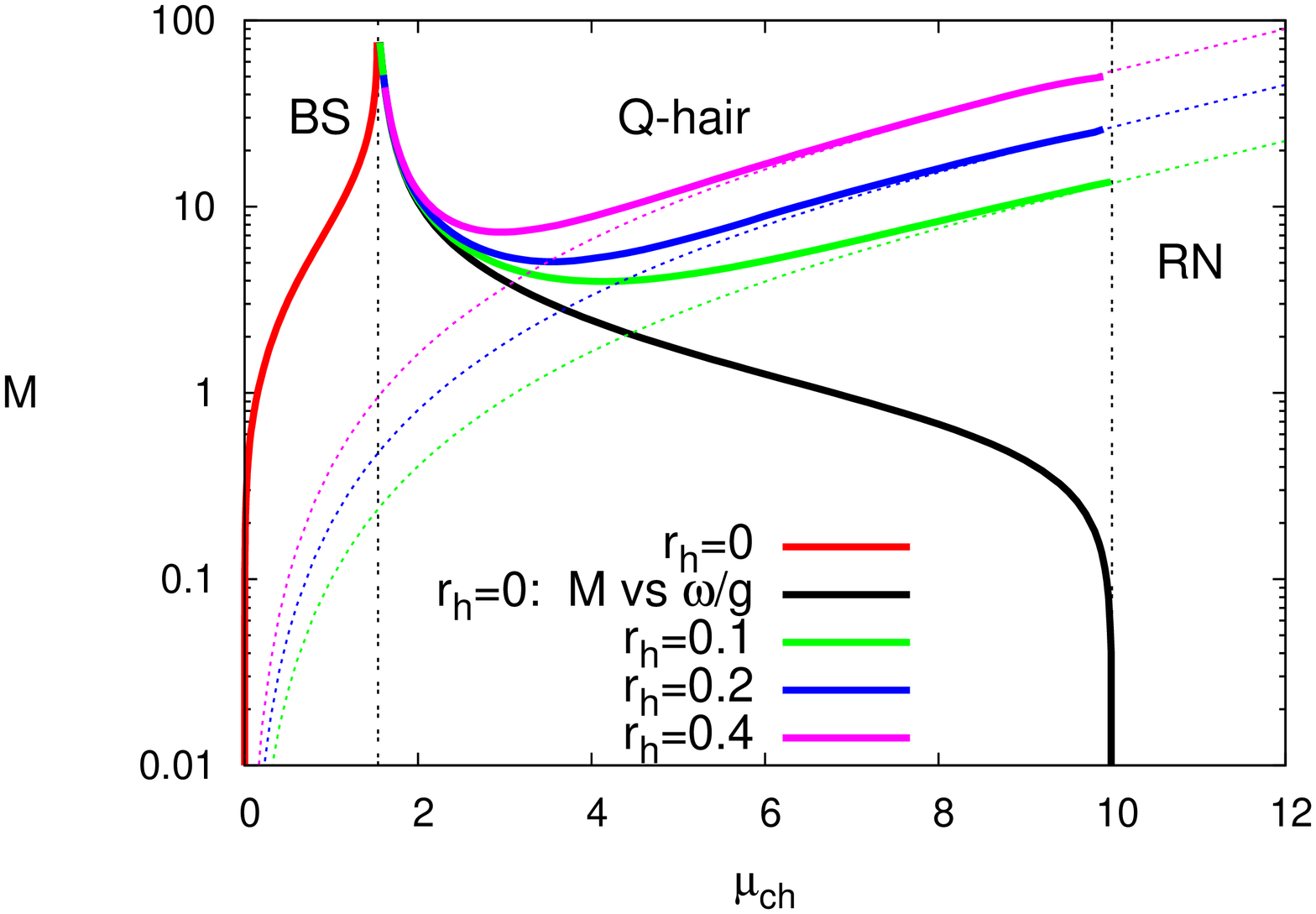}
\includegraphics[height=.33\textheight, angle =-90]{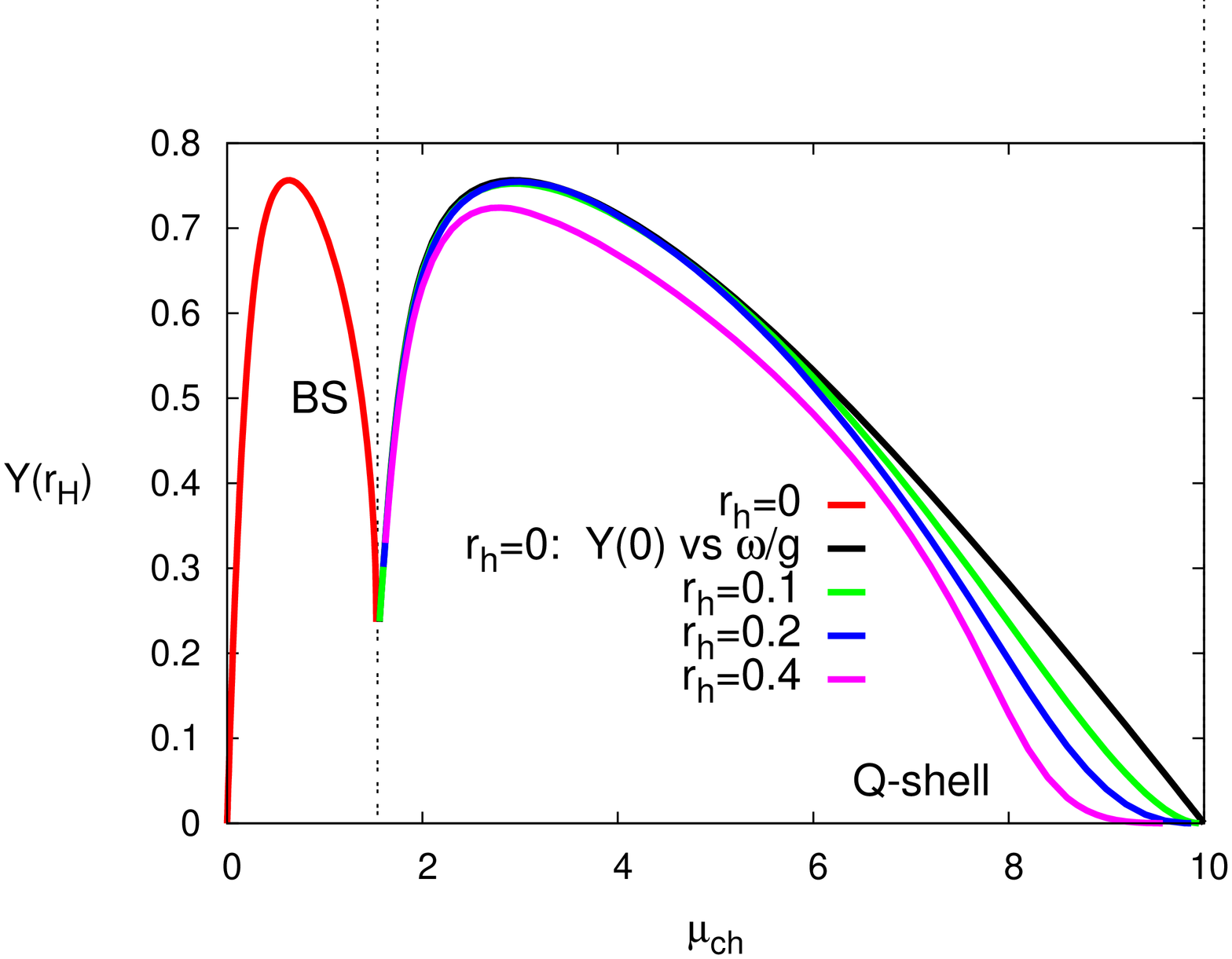}
\end{center}
\caption{\small
Phase structure of charged EMFLS black holes  with resonant Q-hair in the massless ($\mu=0$) limit:
Mass $M$ (left) and horizon value of scalar function $Y(r_H)$ (right) vs chemical potential $\mu_{ch}$ for a set of values of the horizon radius $r_H$, for gravitational coupling $\alpha=0.1$, and gauge coupling $g=0.1$. 
For comparison the corresponding RN properties are shown.
The vertical lines represent the limits of the domain of existence.
$M$ and $Y(r_H)$ are also shown for boson stars versus $\mu_{ch}$ (red) and versus scaled frequency $\omega/g$ (black).
}
    \lbfig{fig5}
\end{figure}

\begin{figure}[h!]
\begin{center}
\includegraphics[height=.33\textheight,  angle =-90]{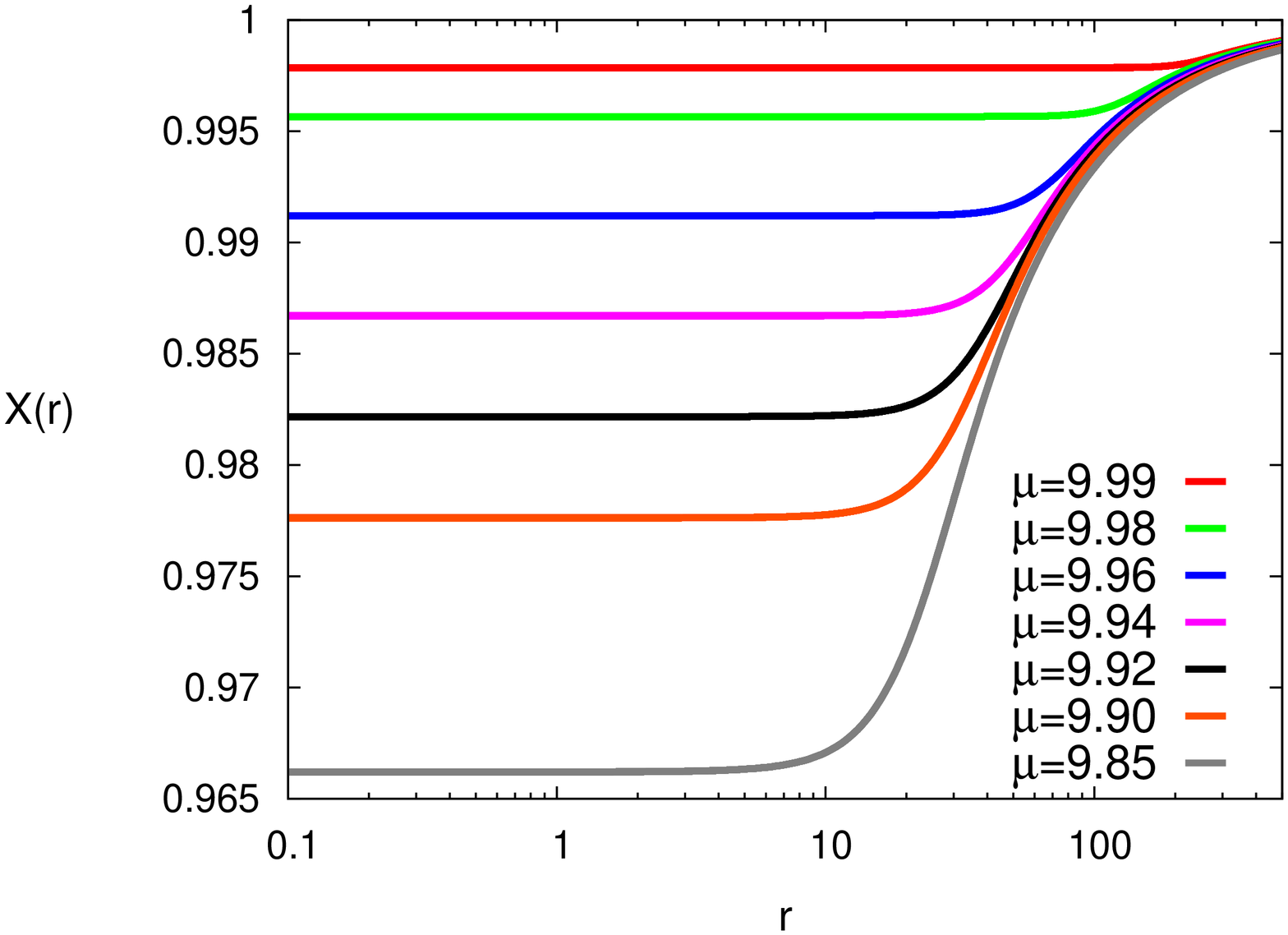}
\includegraphics[height=.33\textheight,  angle =-90]{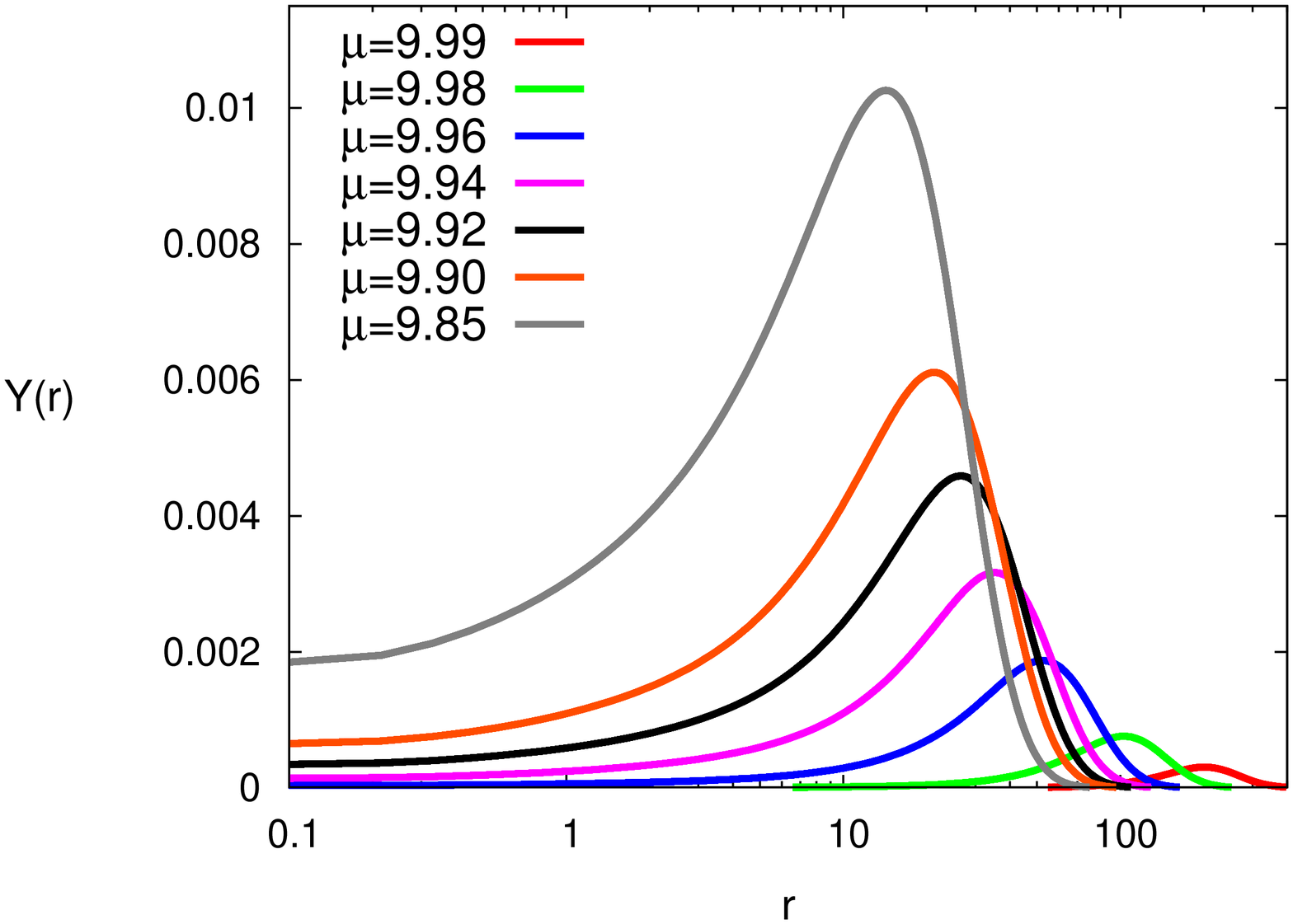}
\end{center}
\caption{\small
Charged EMFLS black holes  with resonant Q-hair in the massless ($\mu=0$) limit:
Profile functions of the scalar fields $X(r)$ (left) and $Y(r)$ (right) for a set of values of the chemical potential $\mu_{ch}$ close to its maximal value, for horizon radius $r_H=0.1$, for gravitational coupling $\alpha=0.1$, and gauge coupling $g=0.1$.
}
    \lbfig{fig6}
\end{figure}

\begin{figure}[tbh]
\begin{center}
\includegraphics[height=.33\textheight,  angle =-90]{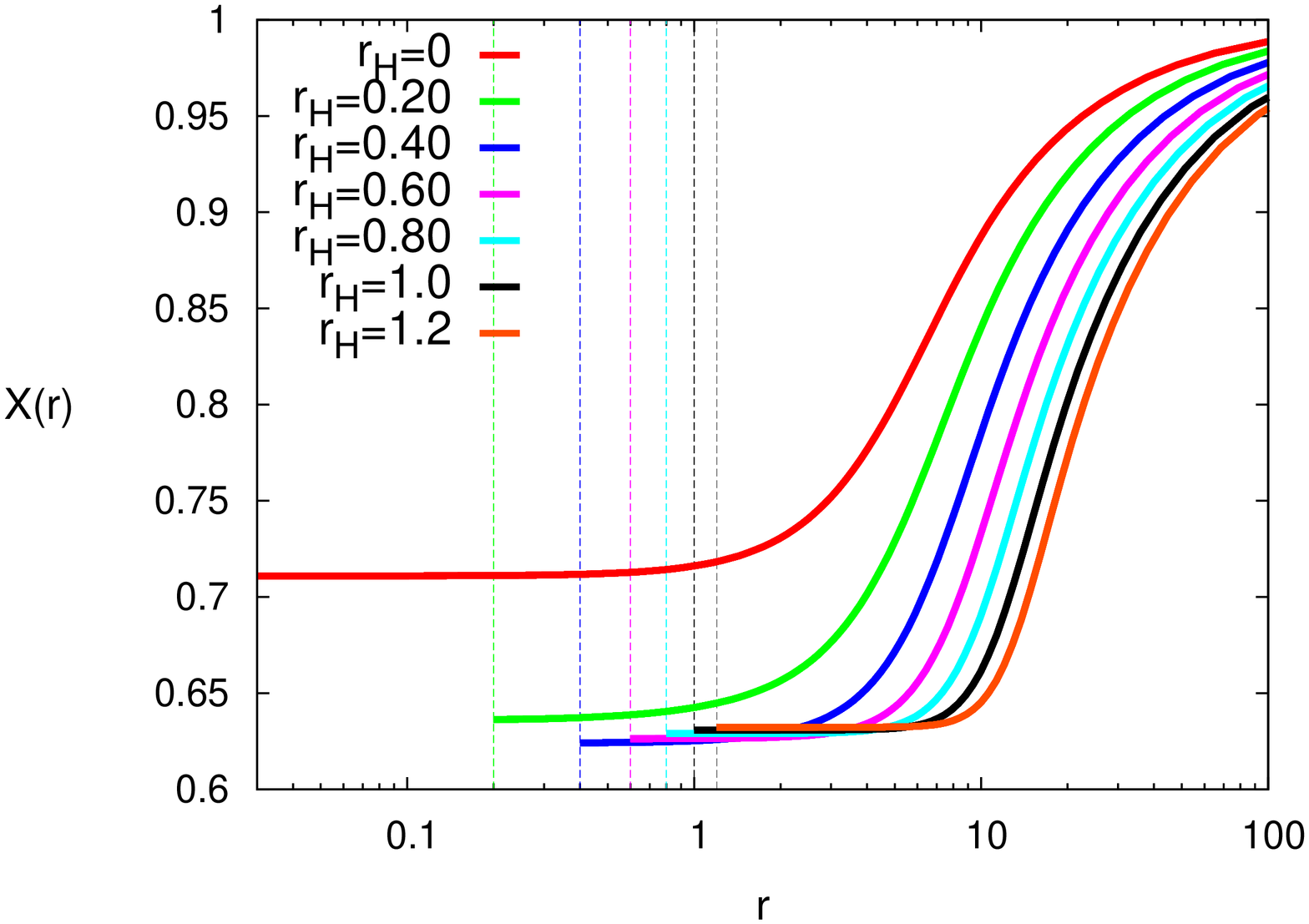}
\includegraphics[height=.33\textheight,  angle =-90]{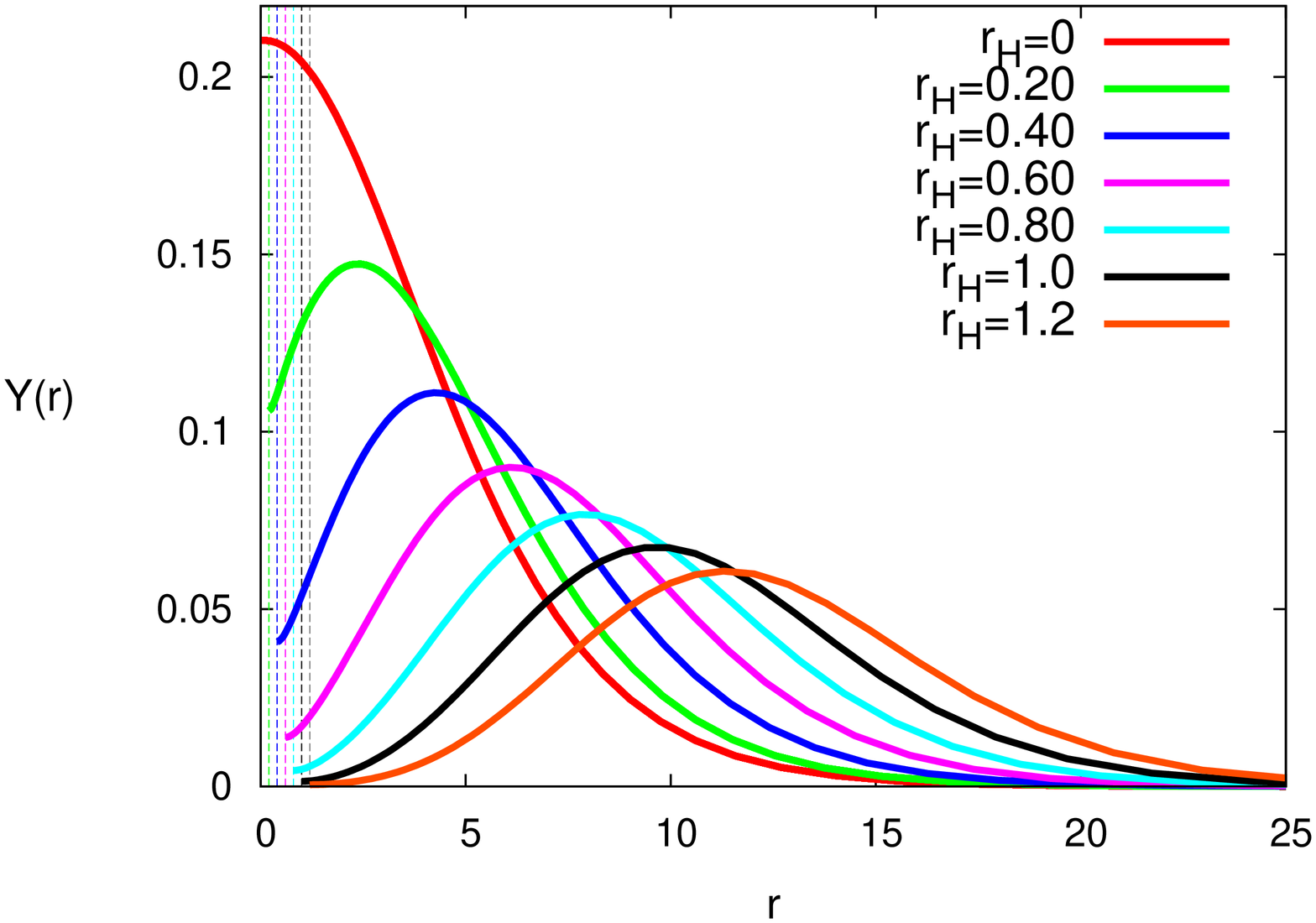}
\includegraphics[height=.33\textheight,  angle =-90]{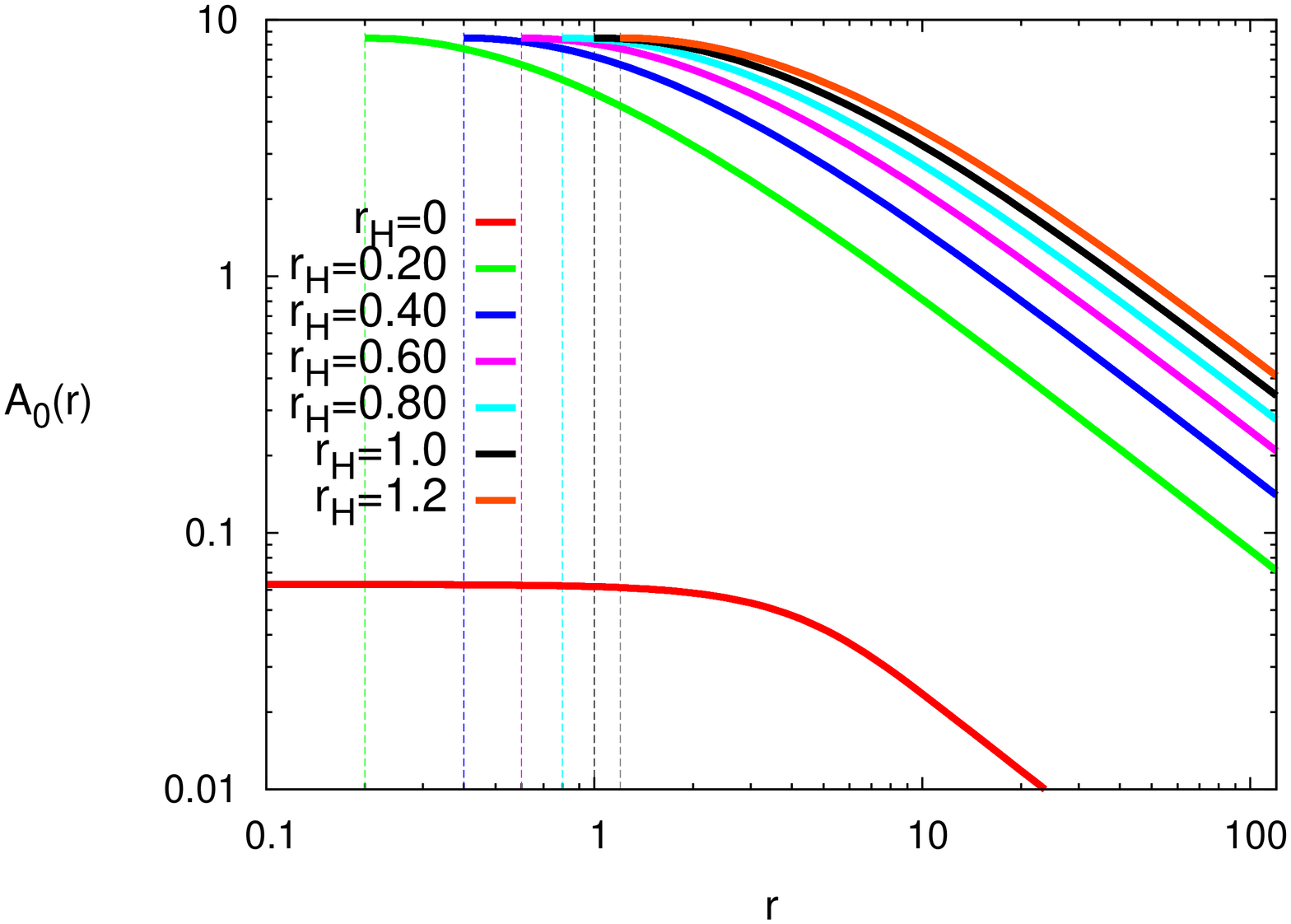}
\includegraphics[height=.33\textheight,  angle =-90]{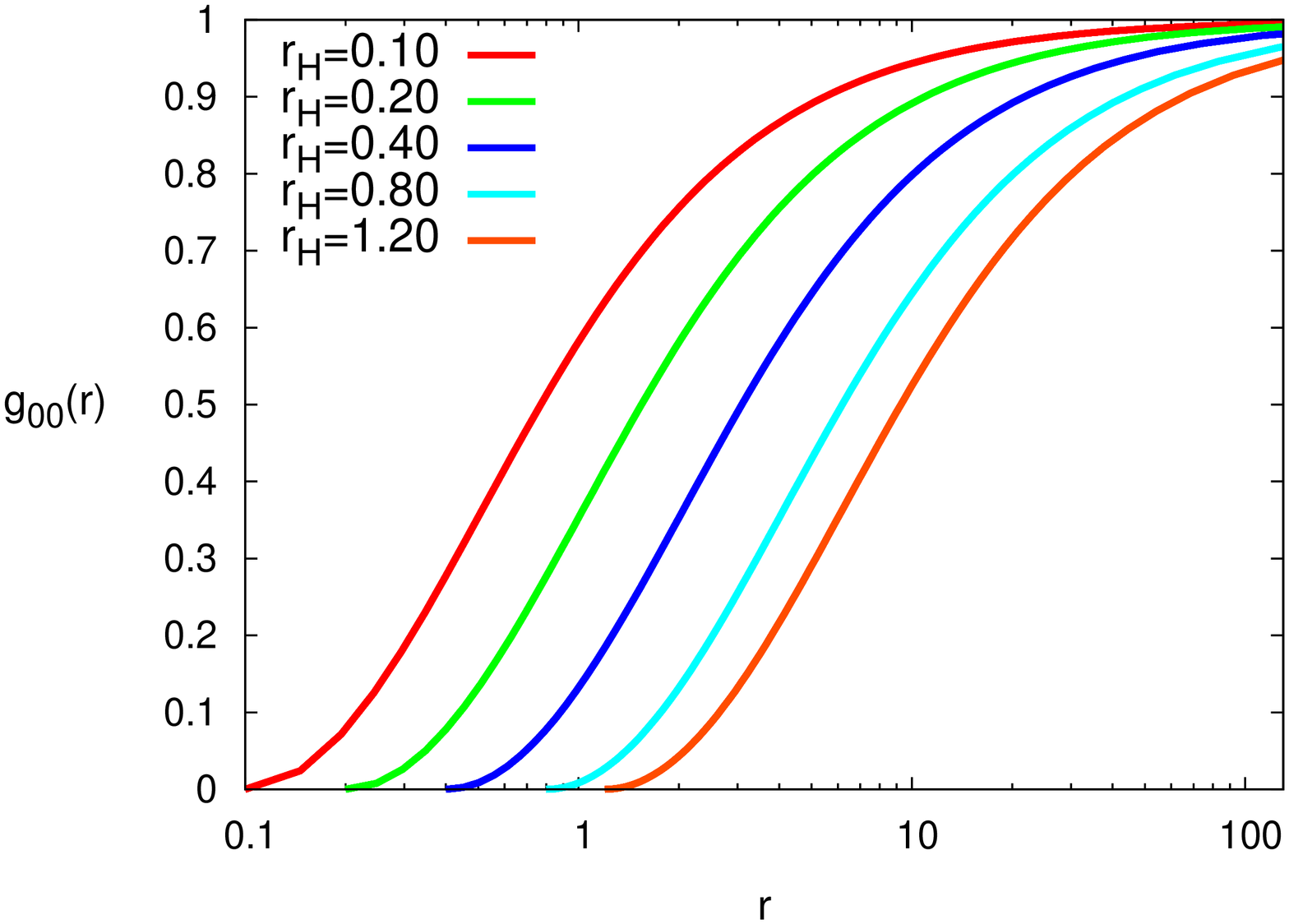}
\end{center}
\caption{\small
Charged EMFLS black holes  with resonant Q-hair in the massless ($\mu=0$) limit:
Profile functions of the scalar fields $X(r)$ (upper left) and $Y(r)$ (upper right), the gauge potential $A_0(r)$ (lower left), and the metric function $g_{00}(r)$ (lower right) for a set of values of the horizon radius $r_H$, for gravitational coupling $\alpha=0.1$, gauge coupling $g=0.1$, and chemical potential $\mu_{ch} =8.5$.}
    \lbfig{fig7}
\end{figure}

\begin{figure}[h!]
\begin{center}
\includegraphics[height=.33\textheight,  angle =-90]{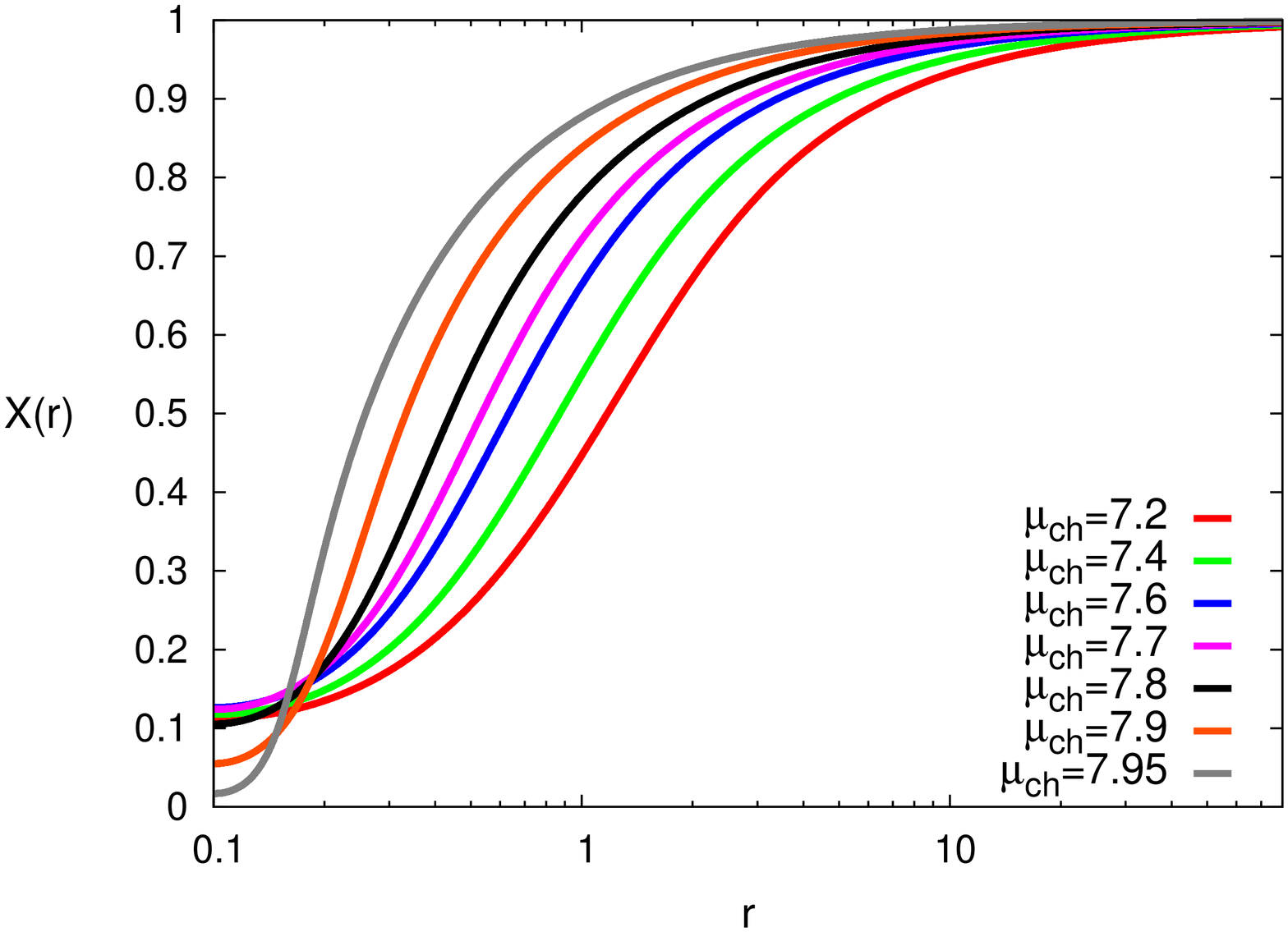} 
\includegraphics[height=.33\textheight,  angle =-90]{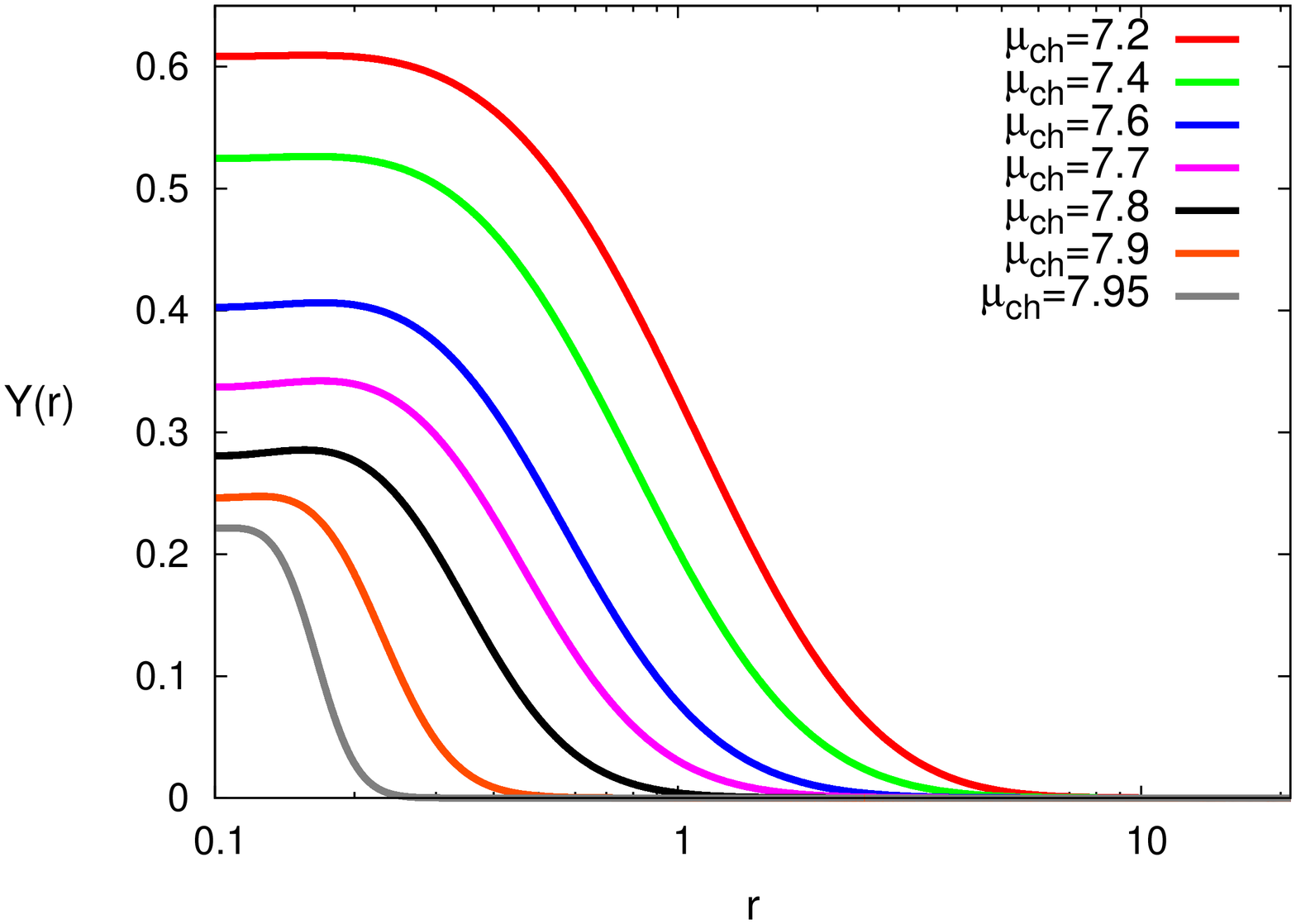}
\end{center}
\caption{\small
Charged EMFLS black holes  with resonant Q-hair in the massless ($\mu=0$) limit:
Profile functions of the scalar fields $X(r)$ (left) and $Y(r)$ (right) for a set of values of the chemical potential $\mu_{ch}$ close to its maximal value, for horizon radius $r_H=0.1$, for gravitational coupling $\alpha=0.25$, and gauge coupling $g=0.1$.
}
    \lbfig{fig8}
\end{figure}

\begin{figure}[h!]
\begin{center}
\includegraphics[height=.33\textheight,  angle =-90]{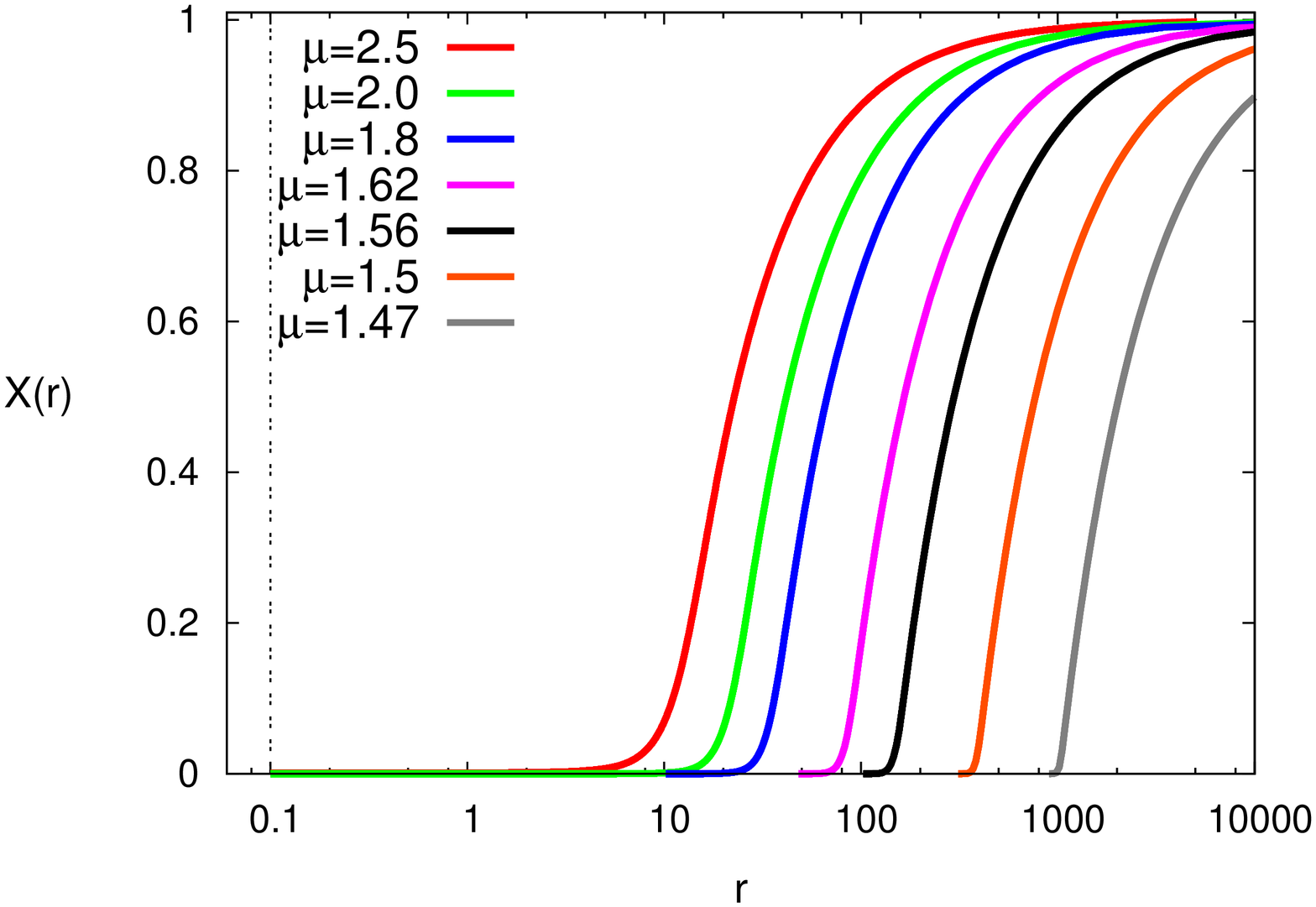}
\includegraphics[height=.33\textheight,  angle =-90]{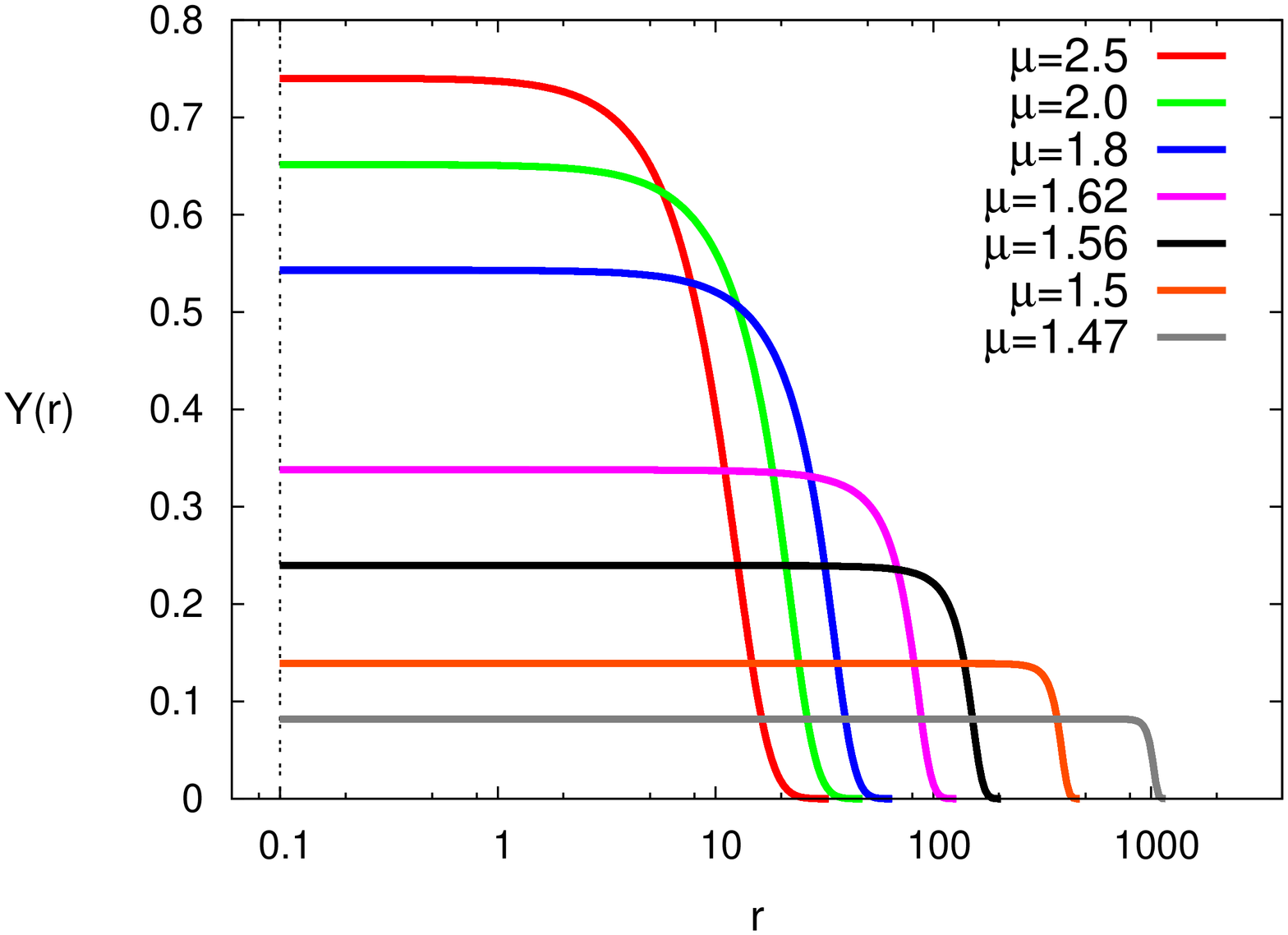}
\includegraphics[height=.33\textheight,  angle =-90]{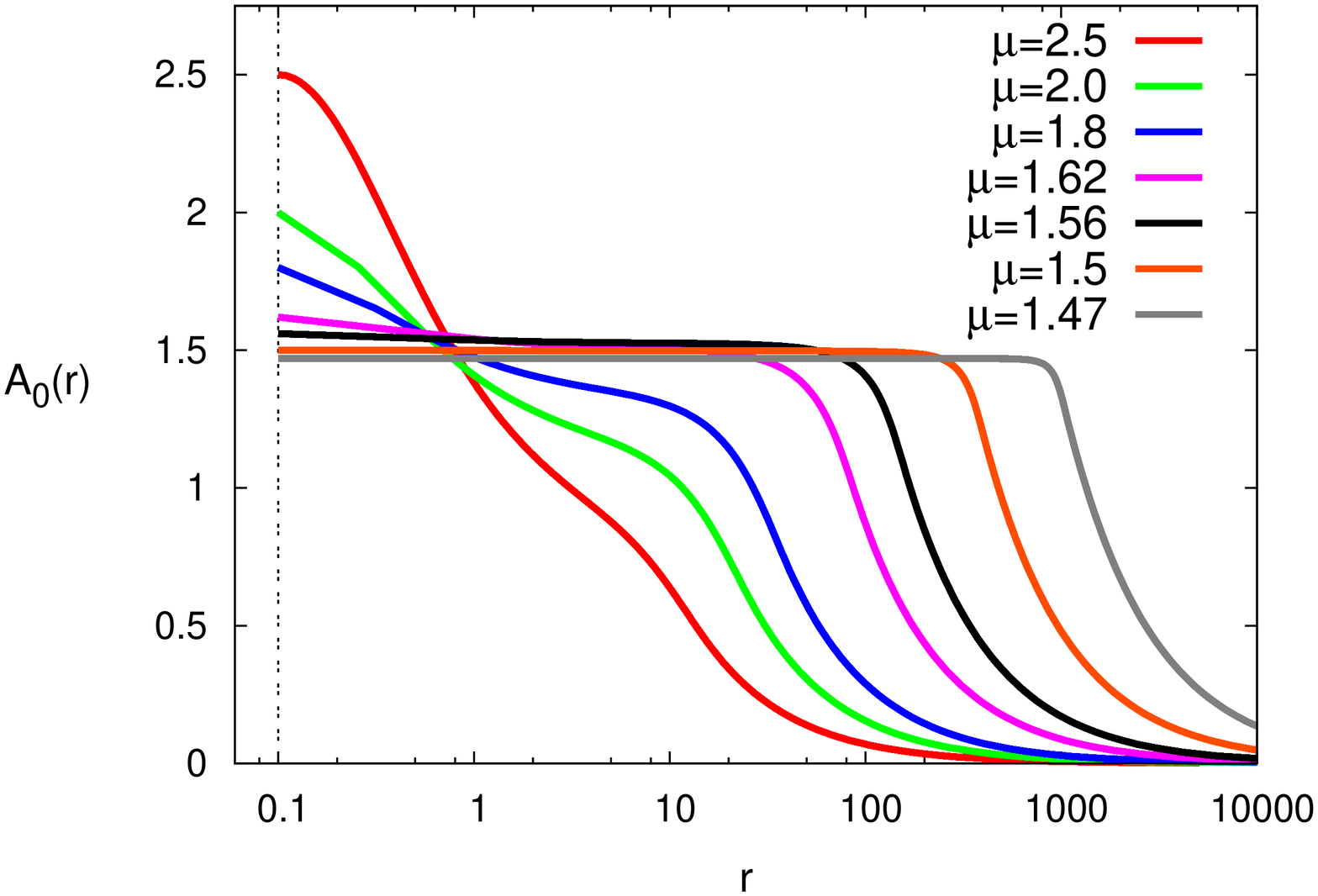}
\includegraphics[height=.33\textheight,  angle =-90]{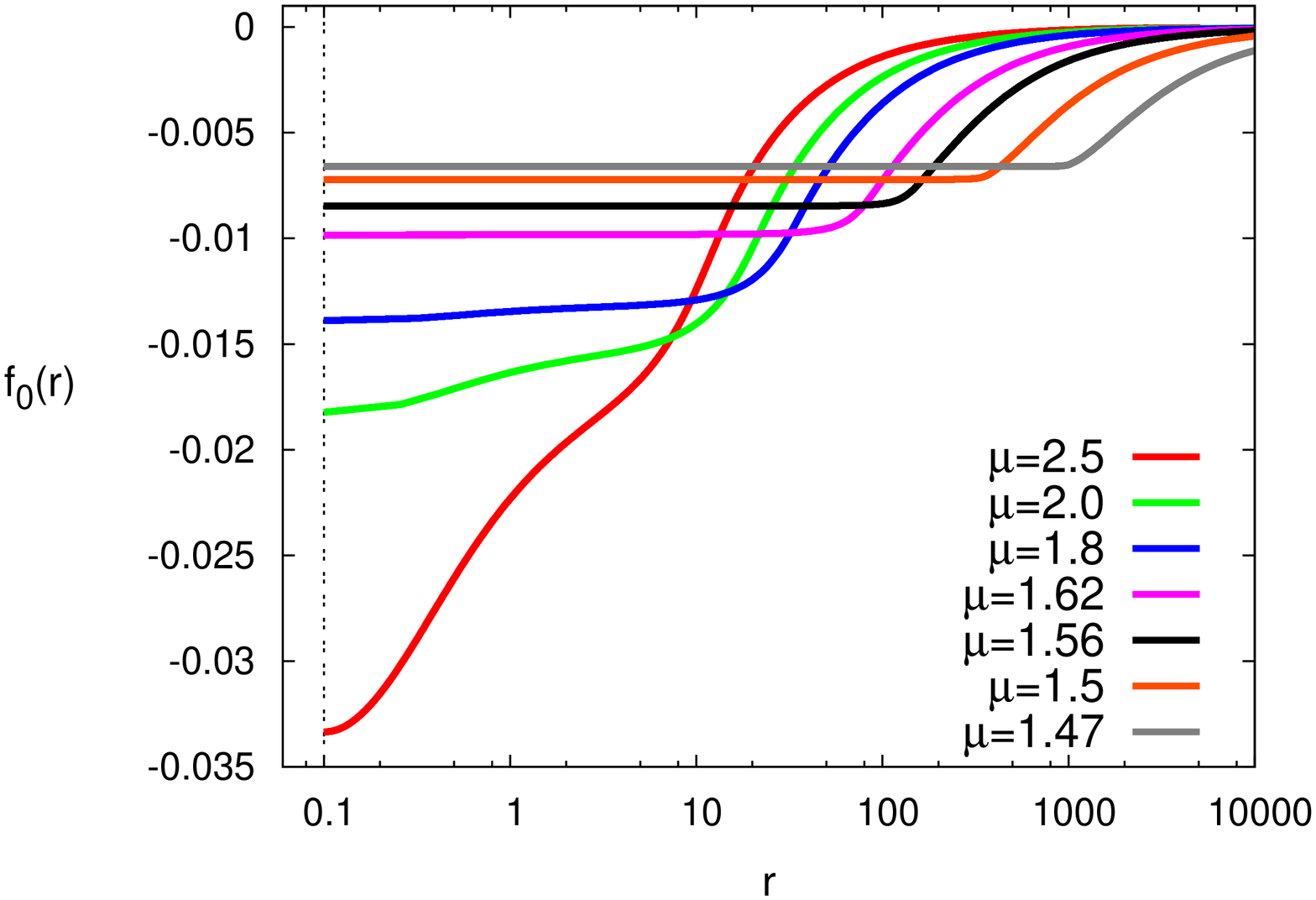}
\end{center}
\caption{\small
Charged EMFLS black holes  with resonant Q-hair in the massless ($\mu=0$) limit:
Profile functions of the scalar fields $X(r)$ (upper left) and $Y(r)$ (upper right), the gauge potential $A_0(r)$ (lower left), and the metric function $f_0(r)$ (lower right) for a set of values of the chemical potential $\mu_{ch}$, for horizon radius $r_H=0.1$, for gravitational coupling $\alpha=0.15$, and gauge coupling $g=0.1$.
}
\lbfig{fig9}
\end{figure}

The above considerations have shown that, as the mass parameter $\mu$ remains finite, the properties of the $Q$ hairy black holes depend crucially on the relative strength of the gravitational and electrostatic interactions. 
Moreover, analogous to the corresponding black holes with resonant single-component scalar Q-hair \cite{Herdeiro:2020xmb,Hong:2020miv}, there is always a gap between the hairy black holes and the Reissner-Nordstr{\"{o}}m black holes (see Figs.~\ref{fig1}-\ref{fig3}).

However, the situation radically changes in the massless limit $\mu = 0$ as the real component of the scalar field becomes long-ranged. 
We recall that in this case there is only a single branch of regular EMFLS Q-balls in flat space, starting from the maximal critical value of the angular frequency and extending all the way down to the limit $\omega\to 0$ \cite{Loiko:2018mhb}.
For boson stars at sufficiently low gravitational coupling we also observe only a single branch which, however, terminates at a finite minimal value of the angular frequency \cite{Kunz:2021mbm}.
Certainly, a small RN black hole can be immersed in the interior region of a gauged EMFLS boson star also in the massless limit.
But we expect a significant change of the pattern displayed by the Q-hairy black holes when $\mu=0$.

To demonstrate the new pattern, we first keep the horizon radius and the gauge coupling fixed, and consider several values of the gravitational coupling $\alpha$. 
We then keep the gravitational coupling $\alpha$ and the gauge coupling fixed, and consider several values of the horizon radius.
We display in Figs.~\ref{fig4} and \ref{fig5} the main characteristics of the resulting Q-hairy black holes in the massless limit.
For comparison we also include the corresponding properties of the boson stars and the RN black holes, when suitable.

Surprisingly, we observe that for large values of the chemical potential $\mu_{ch}$ the mass $M$ and the charge $Q$ of the Q-hairy black holes agree closely with the corresponding values of the RN black holes (see Fig.~\ref{fig4} (upper left, lower left) and Fig.~\ref{fig5} (left)).
Also the hairiness (Fig.~\ref{fig4} (lower right)) seems to start from $h=0$, indicating the possibility of a linear emergence of the Q-hairy black holes from the RN solutions.

When contemplating this possibility, we note that the presence of the long-range scalar field in the massless limit of the two-component EMFLS model may allow to circumvent the no-hair theorem for charged RN black holes due to Hod \cite{Hod:2013nn}. 
Indeed, the corresponding arguments there are based on the presence of a massive charged scalar Klein-Gordon test field on the
fixed RN background, where the mass is a given constant.
However, in the EMFLS model the mass of the complex scalar field is only an effective mass, that possesses a dependence on the radial coordinate for these Q-hairy black hole configurations.
Moreover, the presence of the massless scalar component changes the balance of forces dramatically.

To investigate this intriguing possibility, we first consider relatively small gravitational coupling, $\alpha \le 0.17$.
Here we observe that, as $\alpha$ remains relatively weak, the branch of Q-hairy black holes indeed emerges smoothly from small fluctuations of the charged scalar field in the background of a RN black hole at the maximal value of the chemical potential $\mu_{ch}$.
However, the horizon value of the profile function $Y(r_H)$, shown in Fig.~\ref{fig4} (middle right) versus the chemical potential $\mu_{ch}$, indicates that instead of an expected linear cloud a small linear spherical charged Q-shell is formed around the event horizon.

This is demonstrated in Fig.~\ref{fig6}, where we show for the Q-hairy black holes the change of the real scalar field (left) and the complex scalar field (right), as the maximal value of the chemical potential $\mu_{ch}=10$ is approached.
The electric charge of the complex scalar field is localized inside the spherical shell, which is stabilized by the force balance between the gravitational attraction and the electrostatic repulsion. 
Thus the configuration possesses a region without the charged complex matter field between the horizon of the black hole and the spherical electrically charged Q-shell.

Figure \ref{fig7} illustrates the profile functions of the scalar fields $X(r)$ (upper left) and $Y(r)$ (upper right), the gauge potential $A_0(r)$ (lower left), and the metric function $g_{00}(r)$ (lower right) for a set of values of the horizon radius $r_H$, keeping the other parameters fixed.
The choice for gravitational coupling $\alpha=0.1$ allows to illustrate the transition from Q-hairy black holes with Q-clouds to Q-hairy black holes with Q-shells.
An increase of the horizon area $A_H$, related with a corresponding increase of the horizon charge $Q_H$, causes a stronger repelling force between the horizon and the Q-cloud.
Consequently, at a critical value of the horizon area a Q-shell is formed, as seen in  Fig.~\ref{fig7} (upper right). 
More generally, an increase of the horizon area drives the Q-hairy  black holes towards the limiting RN solution surrounded by a Q-shell.

We note that Q-shell solutions in flat space are known to occur in various models, typically with a V-shape potential \cite{Arodz:2008nm,Ishihara:2021iag,Klimas:2017eft,Heeck:2021bce}.  
In such a case the electrostatic repulsion is balanced by a repulsive interaction mediated by the matter fields. 
Unlike Q-balls, Q-shell solutions in flat space do not form branches.
They may exist only for a given set of parameters of the model. 
Analogous Q-shell configurations are also known in several gauged models coupled to gravity, where they may also harbor a black hole \cite{Kleihaus:2009kr,Kleihaus:2010ep,Hartmann:2013kna,Kumar:2014kna,Klimas:2018ywv,Sawado:2020ncc}.

When the gravitational coupling $\alpha$ increases the distance between the horizon and the shell decreases.
As $\alpha$ increases above $\alpha \sim 0.2$, gravity becomes sufficiently strong to dominate over the electrostatic repulsion.
For the large values of $\alpha$ the branches of Q-hairy black holes no longer start from $\mu_{ch}=10$.
Instead they exhibit smaller maximal values of the chemical potential, that decrease with increasing $\alpha$.
While this may seem surprising, it is in fact again a consequence of our choice of isotropic coordinates and fixed small isotropic horizon radius.
RN black holes then possess a limiting value for the chemical potential, and mass and charge increase to infinity, as this limiting value is approached.

To understand the corresponding behavior of the Q-hairy black holes we inspect Fig.~\ref{fig8}.
The figure shows, that as the limiting chemical potential $\mu_{ch}$ is approached the complex scalar field contracts more and more towards the horizon and at the same time decreases in amplitude and size.
In contrast the real scalar is very small in vicinity of the horizon and then rises more and more steeply towards its vacuum expectation value.
This limiting behavior shows that for large $\alpha$ the Q-hairy black holes do not arise from small fluctuations around RN black holes.

Turning back to Fig.~\ref{fig4} we now consider the properties of the Q-hairy black holes for smaller values of the chemical potential.
As the chemical potential decreases below $\mu_{ch}^{cr}$, the electrostatic interaction begins to overshadow gravity and the Q-cloud around the horizon rapidly inflates. 
This leads to the second, electrostatic branch of solutions, which follows the evolution pattern of the corresponding EMFLS boson stars. 
Both the mass $M$ and the charge $Q$ of the configurations then diverge at a minimal value of the chemical potential $\mu_{ch}$, where the hairiness parameter $h$ approaches unity. 
The critical minimal value of the chemical potential increases with increasing gravitational coupling $\alpha$. 

Remarkably, this minimal value of the chemical potential of the Q-hairy black holes corresponds to the maximal value of the chemical potential allowed for the regular EMFLS boson stars, as seen in Fig.~\ref{fig5} (left). 
In other words, there is no fully continuous transition between Q-hairy EMFLS black holes and regular boson stars, since several horizon values are not continuous in the limit $r_H \to 0$.
Global quantities on the other hand do change continuously in the limit.
This is seen in the figure for the mass, where also the scaled rotational frequency $\omega/g$ is employed for the EMFLS boson stars (black).

Towards the minimal value of the chemical potential we observe remarkably little dependence of the Q-hairy black holes on the gravitational coupling $\alpha$.
In fact all curves in Fig.~\ref{fig4} are getting very close as the lower limiting value of the chemical potential $\mu_{ch}$ is approached.
For gravitational coupling $\alpha=0.15$, horizon radius $r_H=0.1$, and gauge coupling $g=0.1$ the profile functions of the scalar fields $X(r)$ (upper left) and $Y(r)$ (upper right), the gauge potential $A_0(r)$ (lower left), and the metric function $f_0(r)$ (lower right) are exhibited in Fig.~\ref{fig9} for a set of values of the chemical potential $\mu_{ch}$ approaching the minimal value. 
Thus the figure illustrates clearly the rapidly inflating Q-cloud around the horizon, while the contribution of the horizon becomes increasingly less relevant.

\section{Conclusion}

In this work we have considered spherically symmetric gauged black holes with synchronised charged scalar hair in the two-component renormalizable Einstein-Maxwell-Friedberg-Lee-Sirlin model.
This theory may serve as a toy-model to study matter fields localized by gravity in more realistic theories with symmetry breaking potential, like the Standard Model. 
We have shown that the distinctive new features of the Q-hairy EMFLS black holes are related to the delicate force balance between gravitational attraction, electrostatic repulsion between the horizon of the black hole and Q-clouds, as well as scalar interactions, which may also be long-ranged. 

The most remarkable new feature occurs indeed in the case of a massless real scalar field ($\mu=0$).
Here our results indicate that for sufficiently small gravitational coupling Q-hairy black holes arise at the maximal value of the chemical potential (set by the ratio of the mass $m$ of the complex scalar field and the gauge coupling $g$) from small fluctuations around Reissner-Nordstr\"om black holes.
A rigorous proof for the massless case of the EMFLS model would certainly be desirable, following possibly the analysis given by Hod for the massive Klein-Gordon linearized system \cite{Hod:2012wmy,Hod:2013nn}.

Thus, in the massless limit and for sufficiently small gravitational coupling Q-hairy EMFLS black holes can be smoothly linked to the Reissner-Nordstr{\"{o}}m black holes.
This result is contrary to the previously known charged black holes with resonant Q-hair in the Einstein-Maxwell-scalar models with a single scalar field \cite{Herdeiro:2020xmb,Hong:2020miv,Brihaye:2021phs} where a gap is always present. 
Moreover, as the Q-hairy black holes arise from the Reissner-Nordstr\"om black holes linear Q-shells are formed around the black holes instead of (possibly expected) linear Q-clouds.

While the global charges of the Q-hairy black holes seem to suggest, that the massless real scalar field might also allow the Q-hairy black holes to be smoothly linked to the Reissner-Nordstr{\"{o}}m black holes at larger gravitational coupling, closer inspection of the behavior of the functions does not support this.
For larger coupling a very different limiting behavior is observed, where the Q-cloud around the horizon decreases in amplitude and size while the real scalar field becomes very small in the vicinity of the horizon, and then rises sharply to its vacuum expectation value. 
At the same time the metric functions resemble closely those of the Reissner-Nordstr\"om black holes, showing that for the corresponding large values of the global charges the contributions of the scalar fields become basically negligible.

While we have restricted our discussion here to the relatively simple spherically-symmetric system, there remains the
task to consider axially-symmetric gauged Q-hairy EMFLS black holes which may possess both electric charge and angular momentum. 
Here the resulting presence of a magnetic field should provide new interesting features.
It should also be interesting to construct Q-hairy EMFLS black holes in asymptotically AdS spacetime and investigate their possible interpretation in an AdS/CFT context.

\section*{Acknowledgment}
This work was supported by the DFG project Ku612/18-1, the Alexander von Humboldt Foundation, the Heisenberg-Landau program as well as the DAAD Ostpartnerschaftsprogramm.

 \begin{small}

\end{small}

\end{document}